\documentclass[aps,prl,floatfix,twocolumn,superscriptaddress]{revtex4-1}
\usepackage{amsmath,amssymb}
\usepackage{graphicx}
\usepackage{epsfig}
\usepackage{psfrag}
\usepackage[usenames]{color}
\usepackage{xcolor}
\usepackage{hyperref}
\usepackage{soul,color}
\usepackage{physics}
\usepackage{tabu}
\usepackage{multirow}
\hypersetup{colorlinks=true,citecolor={blue},linkcolor={blue},urlcolor={blue}}

\begin{document}





\title{Unconventional  Bloch-Gr\"{u}neisen scattering in hybrid Bose-Fermi systems}



\author{K.~H.~A.~Villegas}
\affiliation{Center for Theoretical Physics of Complex Systems, Institute for Basic Science (IBS), Daejeon 34126, Korea}

\author{Meng~Sun}
\affiliation{Center for Theoretical Physics of Complex Systems, Institute for Basic Science (IBS), Daejeon 34126, Korea}
\affiliation{Basic Science Program, Korea University of Science and Technology (UST), Daejeon 34113, Korea}

\author{V.~M.~Kovalev}
\affiliation{A.~V.~Rzhanov Institute of Semiconductor Physics, Siberian Branch of Russian Academy of Sciences, Novosibirsk 630090, Russia}
\affiliation{Department of Applied and Theoretical Physics, Novosibirsk State Technical University, Novosibirsk 630073, Russia}

\author{I.~G.~Savenko}
\affiliation{Center for Theoretical Physics of Complex Systems, Institute for Basic Science (IBS), Daejeon 34126, Korea}
\affiliation{Basic Science Program, Korea University of Science and Technology (UST), Daejeon 34113, Korea}
\affiliation{A.~V.~Rzhanov Institute of Semiconductor Physics, Siberian Branch of Russian Academy of Sciences, Novosibirsk 630090, Russia}

\date{\today}

\begin{abstract}
We report on the novel mechanism of electron scattering in hybrid Bose-Fermi systems consisting of a two-dimensional electron gas in the vicinity of an exciton condensate: 
We show that a pair-of-bogolons--mediated scattering proves to be dominating over the conventional acoustic phonon channel and over the single-bogolon scattering, even if the screening is taken into account. We develop a microscopic theory of this effect, focusing on GaAs and MoS$_2$ materials, and find the principal temperature dependence of resistivity, distinct from the conventional phonon--mediated processes. 
Further, we scrutinize parameters and suggest a way to design composite samples with predefined electron mobilities and propose a mechanism of electron pairing for superconductivity.

\end{abstract}	

\maketitle


Hybrid Bose-Fermi systems essentially represent a layer of fermions, usually two-dimensional electron gas (2DEG), coupled to another layer of bosons, such as excitons, exciton polaritons, or Cooper pairs.
The interplay between Bose and Fermi particles leads to various novel fascinating phenomena, interesting from both the technological and fundamental physics perspectives. 
For instance, in a hybrid two-dimensional electron gas--superconductor system it became possible to realize the long-sought Majorana fermion~\cite{Sau2010, Alicea2010, Mourik2012, Suominem2017}.
There were also proposed new mechanisms of electron pairing~\cite{Laussy:2010aa} in a hybrid setup involving exciton polaritons in a semiconductor microcavity, opening a possibility for optically controlled superconductivity~\cite{Cotleifmmode-telse-tfi:2016aa}. Furthermore, the interplay between the polaritons and phonons can enhance the critical temperature of the superconductor~\cite{Skopelitis:2018aa}. These results pave the way for the realization of a high-temperature conventional BCS superconductivity. 

In solid state systems, bosons can undergo a phase transition to a Bose-Einstein condensate (BEC), which has been reported in GaAs~\cite{Kasprzak:2006aa} and MoS$_2$ materials~\cite{Berman2016}. In a hybrid system containing a BEC, there can appear magnetically controlled lasing, the Mott phase transition from an ordered state to electron-hole plasma~\cite{Kochereshko:2016aa}, giant Fano resonances~\cite{Boev:2016aa}, which are also shown to occur for superconductor hybrids~\cite{Villegas:2018aa}, and supersolidity~\cite{Matuszewski:2012aa}.
\begin{figure}[!t]
\includegraphics[width=0.50\textwidth]{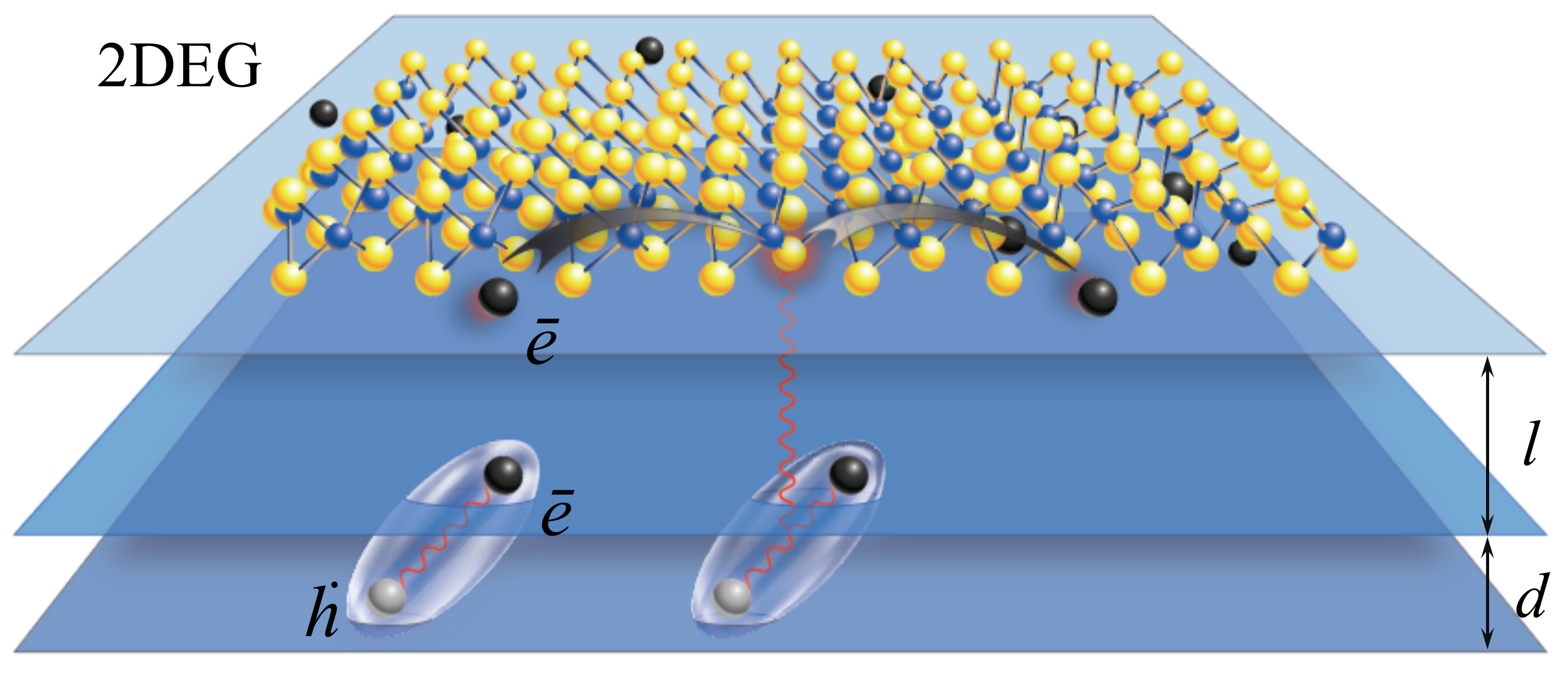}
\caption{System schematic. Bogolon--mediated electron scattering in 2DEG located at the distance $l$ from a two-dimensional dipolar exciton gas, residing in two parallel layers, which are at the distance $d$ from each other. The particles are coupled via the Coulomb interaction.}
\label{fig:1}
\end{figure}
Returning to the fermionic subsystem, studies of the electron transport in 2DEG have many technological applications, especially in the context of interface physics~\cite{Bibes2011, Singhal2003, Sze2007}, where 2DEG exhibits rich phenomena such as the anomalous magnetoresistance and the Hall effect~\cite{Seri2009, Reyren2012, Zhou2015},
two-dimensional metallic conductivity~\cite{Ohtomo2004, Khalsa2012}, superconductivity, and ferromagnetism~\cite{Ganguli2014, Hwang2012, Haraldsen2012, Kumar2015}. %
Electron scattering on acoustic phonons and disorder plays a major role in all these phenomena~\cite{Jena:2007aa,Gibbons:2009aa,Shi:2012aa,Bourgoin:1992aa,Eshchenko:2002aa,Palma:1995aa,Boev:2018ab,Sante2014,Kawamura:1992aa,Gummel:1955aa, Lax:1960aa, Abakumov:1976aa,Kirichenko2017}. 


However, the emerging topic of combining a 2DEG with a BEC demands the study of the electron transport in hybrid systems and forces us to confront new types of interactions beyond the conventional phonon and impurity channels~\cite{Kovalev2011, Kovalev2013, Batyev2014}.
In this Letter, we reexamine the electron transport in hybrid systems and report on the unconventional mechanism of the electron scattering which is due to the interaction with the Bogoliubov excitations or bogolons~\cite{Butov2017,Fogler2014}. The bogolons represent excitations over the BEC and, similar to acoustic phonons, have a linear spectrum at small momenta. While one may naively argue that the bogolon scattering should be similar to the phonon--assisted case, with the acoustic phonon sound velocity simply replaced by the bogolon sound velocity, we will show that this is not at all the case and the difference turns out fundamental. 


Let us consider the system presented in Fig.~\ref{fig:1}, consisting of a 2DEG with parabolic dispersion of electrons and a layer of the Bose-condensed exciton gas~\cite{Butov:2003aa,Kasprzak:2006aa}. The two layers are spatially separated and coupled by the Coulomb interaction~\cite{Boev:2016aa,Kochereshko:2016aa, Matuszewski:2012aa}, 
%
%
%
%
%
described by the Hamiltonian
\begin{equation}\label{eq.1}
V=\int d\mathbf{r}\int d\mathbf{R}\Psi^\dag_\mathbf{r}\Psi_\mathbf{r}g\left(\mathbf{r}-\mathbf{R}\right)\Phi^\dag_\mathbf{R}\Phi_\mathbf{R},
\end{equation}
where $\Psi_\mathbf{r}$ and $\Phi_\mathbf{R}$ are the field operators of electrons and excitons, respectively, $g\left(\mathbf{r}-\mathbf{R}\right)$ is the Coulomb interaction term, $\mathbf{r}$ is the coordinate in the 2DEG plane, and $\mathbf{R}$ is the exciton center-of-mass coordinate. 

Since the excitons are in the BEC phase, we will use the model of weakly interacting Bose gas. Then  $\Phi_\mathbf{R}=\sqrt{n_c}+\phi_\mathbf{R}$, where $n_c$ is the density of particles in the condensate and $\phi_\mathbf{R}$ is the field operator for the bogolons.
%
%
%
\begin{figure}[!b]
\includegraphics[width=0.45\textwidth]{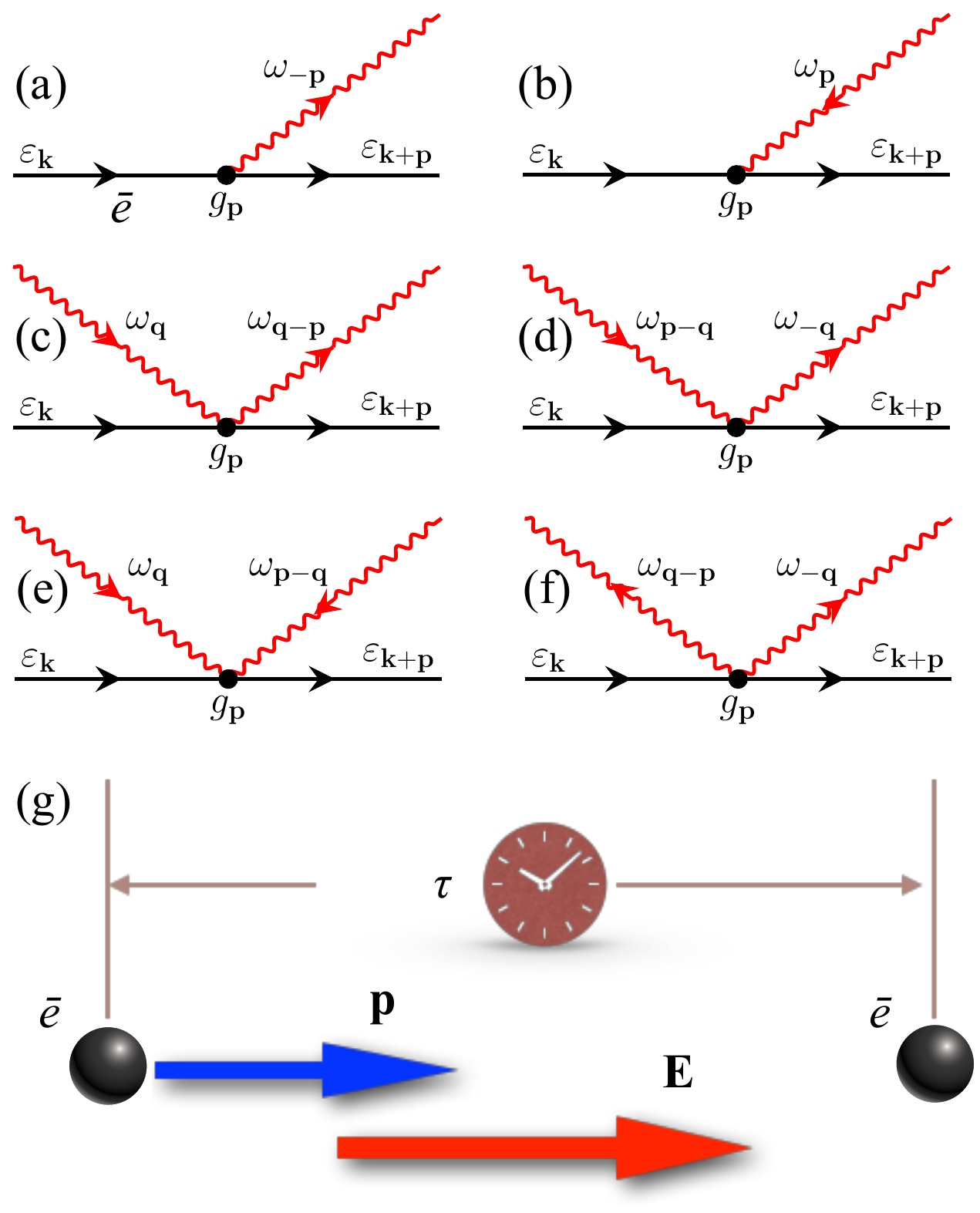}
\caption{Feynman diagrams for the scattering processes: straight black lines represent the electrons, while the wiggly red lines represent the bogolons. (a)-(b) Single-bogolon scattering events. (c)-(f) Two-bogolon scattering. (g) Schematic of the electron distribution function ansatz~\eqref{EqAnsatz} in the Boltzmann equation: the work done by the electric field $\mathbf{E}$ on the electron with momentum $\mathbf{p}$ during the relaxation time $\tau$ changes the electron energy.}
\label{Fig2}
\end{figure}
Then Eq.~\eqref{eq.1} splits into two terms:
\begin{eqnarray}\label{eq.2}
V_1&=&\sqrt{n_c}\int d\mathbf{r}\Psi^\dag_\mathbf{r}\Psi_\mathbf{r} \int d\mathbf{R}g\left(\mathbf{r}-\mathbf{R}\right)\left[\varphi^\dag_\mathbf{R}+\varphi_\mathbf{R}\right],\\\nonumber
V_2&=&\int d\mathbf{r}\Psi^\dag_\mathbf{r}\Psi_\mathbf{r}\int d\mathbf{R}g(\mathbf{r}-\mathbf{R})\varphi^\dag_\mathbf{R}\varphi_\mathbf{R}.
\end{eqnarray}
Furthermore, we express the field operators as the Fourier series
\begin{eqnarray}
\label{eq.3}
\varphi^\dag_\mathbf{R}+\varphi_\mathbf{R}&=&\frac{1}{L}\sum_{\mathbf{p}} e^{i\mathbf{p}\cdot\mathbf{R}} \left[(u_\mathbf{p}+v_{-\mathbf{p}})b_\mathbf{p}+(v_\mathbf{p}+u_{-\mathbf{p}})b^\dag_{-\mathbf{p}}\right],\nonumber\\
\Psi_\mathbf{r}&=&\frac{1}{L}\sum_\mathbf{k}e^{i\mathbf{k}\cdot\mathbf{r}}c_\mathbf{k},\;\;\;\mbox{and}\;\;\;
\Psi_\mathbf{r}^\dagger=\frac{1}{L}\sum_\mathbf{k}e^{-i\mathbf{k}\cdot\mathbf{r}}c_\mathbf{k}^\dagger,
\end{eqnarray}
where $b_{\mathbf{p}}$($c_\mathbf{k}$) and $b^\dag_{\mathbf{p}}$($c^\dagger_\mathbf{k}$) are the bogolon (electron) annihilation and creation operators, respectively, and $L$ is the length of the structure. The Bogoliubov coefficients read~\cite{Giorgini:1998aa}
\begin{eqnarray}\label{eq.4}
u^2_{\mathbf{p}}&=&1+v^2_{\mathbf{p}}=\frac{1}{2}\left(1+\left[1+\frac{(Ms^2)^2}{\omega^2_{\mathbf{p}}}\right]^{1/2}\right),\\\nonumber
&&u_{\mathbf{p}}v_{\mathbf{p}}=-\frac{Ms^2}{2\omega_{\mathbf{p}}},
\end{eqnarray}
where $M$ is the exciton mass, $s=\sqrt{\kappa n_c/M}$ is the sound velocity, $\kappa=e_0^2d/\epsilon$ is the exciton--exciton interaction strength in the reciprocal space, $e_0$ is electron charge, $\epsilon$ is the dielectric function,
$\omega_k=sk(1+k^2\xi^2)^{1/2}$ is the spectrum of bogolons, and $\xi=\hbar/(2Ms)$ is the healing length. %
Combining Eqs.~\eqref{eq.2} and~\eqref{eq.3}, we find
\begin{eqnarray}
\label{v1}
V_1 &=& \frac{\sqrt{n_c}}{L} \sum_{\mathbf{k,p}} g_\mathbf{p} \left[ \left( v_\mathbf{p} + u_\mathbf{-p} \right)b^\dagger_\mathbf{-p} \right.\\
\nonumber
&&\left.~~~~~~~~~~~~~~~~+ 
\left( u_\mathbf{p} + v_\mathbf{-p}\right)b_\mathbf{p} \right] c^\dagger_\mathbf{k+p} c_\mathbf{k}, \label{eq.4.1}
\\
\label{v2}
V_2 &=& \frac{1}{L^2}\sum_{\mathbf{k,p,q}}g_\mathbf{p}  
\left( u_\mathbf{q-p}u_\mathbf{q} b^\dagger_\mathbf{q-p}b_\mathbf{q} + u_\mathbf{q-p}v_\mathbf{q}b^\dagger_\mathbf{q-p}b^\dagger_\mathbf{-q}  \right.\\
\nonumber
&&\left.+ v_\mathbf{q-p}u_\mathbf{q}b_\mathbf{-q+p}b_\mathbf{q}   +v_\mathbf{q-p}v_\mathbf{q} b_\mathbf{-q+p} b^\dagger_\mathbf{-q}\right)  c^\dagger_\mathbf{k+p} c_\mathbf{k},
\label{eq.4.2}
\end{eqnarray}
%
%
%
where $g_\mathbf{p}=2\pi e^2_0 (1-e^{-pd})e^{-pl}/(\epsilon p)$ is the Fourier image of the electron-exciton interaction. 
Equations~\eqref{v1} and~\eqref{v2} give matrix elements of electron scattering in two conceptually different processes within the same (first) order with respect to the interaction strength $g_\mathbf{p}$. The contribution $V_1$ is responsible for the electron scattering with emission/absorption of a single Bogoliubov quantum, whereas $V_2$ describes the electron scattering mediated by the emission/absorption of a pair of bogolons, which we will refer to as \textit{two-bogolon processes}.


To investigate the principal $T$-dependence of single-bogolon resistivity at low temperatures, we will adopt the Bloch-Gr\"uneisen formalism~\cite{Ziman:2001aa, Zaitsev:2014aa}, which was originally used to describe electron-phonon interaction.
We start from the Boltzmann equation
\begin{eqnarray}
\label{Boltzmann}
e_0\textbf{E}\cdot\frac{\partial f}{\hbar\partial \textbf{p}}=I\{f\},
\end{eqnarray}
where $f$ is the electron distribution, $\mathbf{p}$ is the wave vector, $\mathbf{E}$ is the perturbing electric field, and $I\{f\}$ is the collision integral involving single-bogolon scattering processes, as shown in Fig.~\ref{Fig2} (a) and (b) (see Appendix A in the Supplemental Material ~\cite{[{See Supplemental Material at [URL] for the detailed  derivations}]SMBG} for the explicit form of $I$ and other details of derivation).
For relatively weak perturbing electric fields, $f$ can be expanded as
\begin{eqnarray}
f=f^0(\varepsilon_p)-\left(-\frac{\partial f^0}{\partial\varepsilon_p}\right)f^{(1)}_\textbf{p},
\end{eqnarray}
where $p\equiv\abs{\mathbf{p}}$, $f^0(\varepsilon_p)$ is the Fermi-Dirac distribution.
The function $f^{(1)}_\textbf{p}$ is the change in energy of the electron due to the applied electric field. Without the loss of generality, we put this electric field to be directed along the $x$-axis and use the ansatz
\begin{equation}
\label{EqAnsatz}
f^{(1)}_\textbf{p}= (e_0E_x)(\hbar m^{-1}p_x)\tau(\varepsilon_p), 
\end{equation}
where $m$ is the effective electron mass in the 2DEG and $\tau(\varepsilon_p)$ is the relaxation time. This ansatz can be understood from Fig.~\ref{Fig2}(g). The factor $e_0E_x$ is the force acting on the electron while $\hbar m^{-1}p_x$ is the electron velocity. The function $f^{(1)}$ therefore gives the work done by the electric field on the electron during the relaxation time $\tau$.
\begin{figure}[!b]
\includegraphics[width=0.498\textwidth]{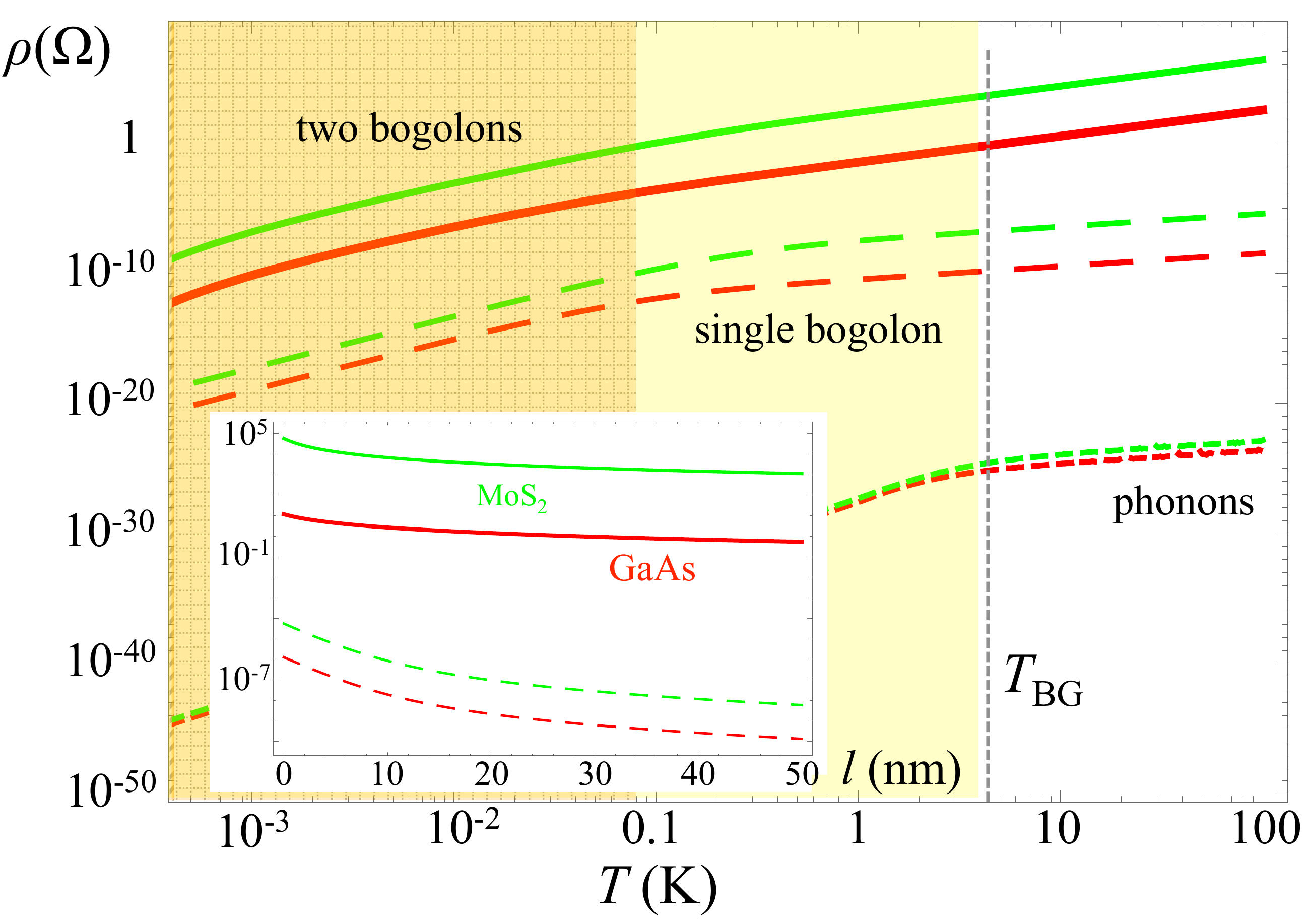}
\caption{Resistivity as a function of temperature (main plot) and layer separation $l$ (inset) with account for the two-bogolon (solid), single-bogolon (dashed), and phonon (dotted curves) scattering. The green curves are for MoS$_2$, while the red curves are for GaAs.}
\label{Fig3}
\end{figure}

After the derivations~\cite{SMBG}, we find the one-bogolon--mediated resistivity, which is the first crucial formula in this Letter:
%
%
%
\begin{eqnarray}
\label{rho1bgen}
\rho^{(1)}=\frac{\pi\hbar^3\xi_I^2}{e_0^2ME_F}\sum_{n=0}^\infty\frac{(-2)^nl^n\gamma_n}{n!(\hbar s)^{n+4}}(k_BT)^{n+4},
\end{eqnarray}
where $\xi_I= e_0^2d\sqrt{n_c}/2\epsilon$, $E_F$ is the Fermi energy, $\gamma_n=(n+3)!\zeta(n+3)/[(2\pi)^2k_BT_\textrm{BG}]$, $T_\textrm{BG}=2\hbar sk_F/k_B$ is the Bloch-Gr\"{u}neisen temperature, $s$ is the sound velocity, $k_F$ is the Fermi wave vector, $k_B$ is the Boltzmann constant, and $\zeta(x)$ is the Riemann zeta function.
The leading term in~\eqref{rho1bgen} at small $T$ reads
\begin{eqnarray}
\label{rho1b}
\rho^{(1)}\approx\frac{\pi\hbar^3\xi_I^2}{e_0^2ME_F}\frac{3!\zeta(3)}{(2\pi)^2k_BT_\textrm{BG}}\left(\frac{k_BT}{\hbar s}\right)^4,
\end{eqnarray}
hence the resistivity behaves as $\rho^{(1)}\propto T^4$ at low temperatures.
%
%
%
%
\begin{figure}[!t]
\includegraphics[width=0.49\textwidth]{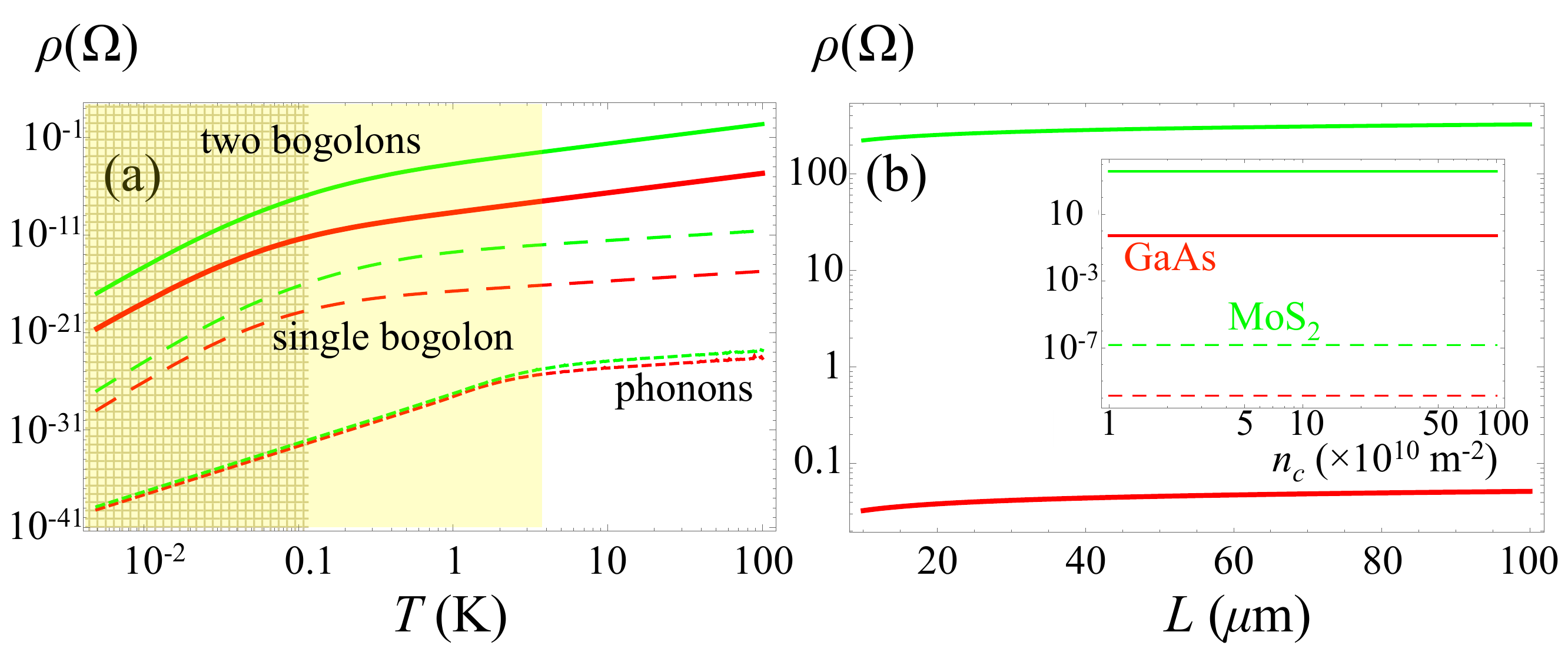}
\caption{Resistivity as a function of temperature in the presence of screening (a) and the sample size for two-bogolon contribution (b). Inset shows the resistivity as a function of the condensate density (see also Fig.~\ref{Fig3} for comparison).}
\label{Fig4}
\end{figure}
%
%
%


The two-bogolon resistivity can also be derived from Eq.~\eqref{Boltzmann}. The collision integral now expresses the net scattering into a state with momentum $\hbar\mathbf{p}$, involving a pair of bogolons, as shown in Fig.~\ref{Fig2}(c)-(f) (see Appendix B in~\cite{SMBG}),
\begin{eqnarray}
\label{rho2bmain}
\rho^{(2)}=\frac{ms^2}{8\pi^2e_0^2mv_F^5}\int\limits_{L^{-1}}^\infty\frac{k^2g_k^2dk}{\sinh^2\left[\frac{\hbar sk}{2k_BT}\right]}\ln(kL),
\end{eqnarray}
where $v_F$ is the Fermi velocity.
This formula is the central result of this Letter. To find~\eqref{rho2bmain}, we used the approximation $v_F\gg s$ and introduced the infrared cut-off $L^{-1}$ for the wave vector integrals, necessary for the convergence. The physical meaning of this cut-off is the absence of fluctuations with he wavelength larger than $L$. This cut-off can also be related to the critical temperature of the Bose-Einstein condensation in a finite trap of length $L$~\cite{Bagnato1991}. Indeed a BEC cannot form in infinite homogeneous 2D systems at finite temperatures~\cite{Hohenberg1967}, thus a trapping with the characteristic size $L$ is required~\cite{Butov2017}.

We can further extract the temperature dependences for the two limits (see Appendix B~\cite{SMBG}). For low temperatures $T\ll T_\textrm{BG}$, we find
\begin{eqnarray}
\label{rho2b}
\rho^{(2)}\approx\frac{s^2e_0^2d^2}{2v_F^5\epsilon^2}\left(\frac{T}{T_\textrm{BG}}\right)^3\frac{\pi}{6(2l)^3}\ln \left(\frac{L}{2l}\right),
\end{eqnarray}
while at high temperatures $T\gg T_\textrm{BG}$,
\begin{eqnarray}
\rho^{(2)}\approx\frac{s^2e_0^2d^2}{2v_F^5\epsilon^2}\left(\frac{T}{T_\textrm{BG}}\right)^2\frac{1}{(2l)^3}\ln \left(\frac{L}{2l}\right).
\end{eqnarray}
%


First of all, we note that at low temperatures, the  single bogolon contribution gives $\rho\sim T^4$ [Eq.~\eqref{rho1b}], while the pair-of-bogolon contribution is $\rho\sim T^3$, which implies that the latter dominates at low temperatures. Figure~\ref{Fig3} shows the temperature behavior of the resistivity contributions from the one-bogolon, two-bogolon, and phonon scattering processes. We used the parameters for GaAs and MoS$_2$ materials~\cite{[{The values of parameters were taken from~\cite{Kaasbjerg2013,Basu1980,Mair1998}. Dielectric constants: $\epsilon_\textrm{GaAs}=12.5\epsilon_0$, $\epsilon_{\textrm{MoS}_2}=4.89\epsilon_0$; effective electron masses ($m_0$ is the bare electron mass): $m_\textrm{GaAs}=0.067m_0$, $m_{\textrm{MoS}_2}=0.47m_0$; exciton masses: $M_\textrm{GaAs}=0.517m_0$, $M_{\textrm{MoS}_2}=0.499m_0$; exciton sizes: $d_\textrm{GaAs}=10.0$ nm, $d_{\textrm{MoS}_2}=3.5$ nm; deformation potentials: $D_\textrm{GaAs}=5.7$ eV, $D_{\textrm{MoS}_2}=2.4$ eV; 
densities of states: DOS$_\textrm{GaAs}=2.13\times 10^{16}$ eV$^{-1}$m$^{-2}$, DOS$_{\textrm{MoS}_2}=4.35\times 10^{18}$ eV$^{-1}$m$^{-2}$; 2D ion mass densities: $3.03\times 10^{-6}$ and $3.1\times10^{-6}$ kg$\cdot$m$^{-2}$ for GaAs and MoS$_2$, respectively}]phonon_para}, where the exciton condensates have been experimentally realized~\cite{Kasprzak:2006aa, Berman2016}.

The main panel of Fig.~\ref{Fig3} shows the resistivity as a function of temperature for typical experimental range $T\lesssim 100$ K at which the exciton condensate exists. The yellow shaded region highlights the Bloch-Gr\"{u}neisen regime $T<T_\textrm{BG}$, where for both GaAs and MoS$_2$ we have $T_\textrm{BG}\approx 5$ K. 
The orange shaded region is the temperature regime, where the resistivity is well approximated by the analytical formulas Eqs.~\eqref{rho1b} and~\eqref{rho2b}.
First, we see that the one- and two-bogolon scattering contributions to the resistivity are orders of magnitude larger than the contribution of the phonon scattering; and second, we see that the two-bogolon scattering processes give a significantly larger contribution to the resistivity than the one-bogolon scattering. 
In the conventional treatment of hybrid 2DEG-BEC systems, the two-bogolon interaction, Eq.~\eqref{eq.4.1}, has been neglected as it was related to the second-order perturbation theory in fluctuations above the macroscopically-occupied ground state. Figure~\ref{Fig3} demonstrates that this widespread approximation is not valid in the context of the indirect exciton condensates. 

The dominance of two-bogolon channel over the  single bogolon scattering can be understood from the analysis of the matrix elements in the Fermi golden rule. In the single bogolon case, there appears a small factor $(u_\mathbf{p}+v_{-\mathbf{p}})\sim\sqrt{1+A^2}-A$, where $A=(Ms)/(\hbar \lambda)$~\cite{SMBG}. In other words, $(u_\mathbf{p}+v_{-\mathbf{p}})\sim(p\xi)^2\ll 1$. In particular, in GaAs and MoS$_2$ materials, this factor is sufficiently small to compensate the large value of $\sqrt{n_c}$. In contrasts, there is no such cancellation effect in the two-bogolon terms, where there appears the product $u_\textbf{p}v_\textbf{p}\sim (p\xi)^{-1}\gg 1$ (instead of $u_\mathbf{p}+v_{-\mathbf{p}}$). Here we would like to draw a comparison with the acoustic phonons, where this cancellation effect does not take place, so that the single-phonon scattering has larger contribution than the two-phonon scattering. This argument manifests the difference between the bogolon and phonon-assisted scattering, which is due to the difference in the origin of interaction. Indeed, the phonon terms appear from the deformation potential theory, while the interaction with bogolons has the Coulomb nature.

Figure~\ref{Fig3}(inset) demonstrates that the resistivity due to any bogolon scattering decreases as the layer separation increases. This is due to the Coulomb interaction between the layers becomes weaker with the distance. This property opens the possibility of designing hybrid Fermi-Bose systems with the desired electron mobility, which is significant in various technological applications. However, $l$ is not the only parameters which might determine the electron scattering. In particular, the dependence on the condensate density $n_c$ and the screening should be addressed.

Figure~\ref{Fig4}(a) shows that the conclusions we made from the analysis of Fig.~\ref{Fig3} still hold even when the screening is taken into account~\cite{SMBG}. The dependence of the resistivity on the sample size $L$ (for two-bogolon scattering, where we introduced the infrared cut-off) and the condensate density $n_c$ are presented in Fig.~\ref{Fig4}(b). The plots show that the resistivity is not sensitive to these parameters. It means that first, we can easily optimize the design by changing $l$ and second, $L$ does not influence the physical phenomena in question. 

In an experiment, it might be difficult to resolve different contributions to the total resistivity. However, using the analytical formulas Eqs.~(\ref{rho1b}) and~(\ref{rho2b}), we see that the low-temperature resistivity should behave as $\sim T^3$, which gives the most considerable contribution. Another obstacle might also arise. At low temperatures, the scattering on disorder starts to play an important role. Thus $T^3$ should be observable in relatively pure samples. Alternatively, high-temperature behavior can be studied, where two-bogolon scattering is $\sim T^2$.

What can we say about the electron pairing in such a hybrid 2DEG-BEC system below $T_c$? For any metal in the normal (not superconducting) state, the strength of electron-phonon interaction is responsible for the resistivity due to the scattering. Obviously, the stronger the interaction strength (which is mostly determined by the matrix element of interaction), the larger is the resistivity. In the superconducting phase, the electron pairing is also mediated by the interaction with phonons (or bogolons~\cite{Laussy:2010aa}). Indeed, there enters the same matrix element of the electron-phonon interaction. The bigger it is, the larger the superconducting gap opens, which means a robust superconductivity. The critical temperature is also determined by the strength of the electron-phonon (bogolon) interaction. It makes us suppose that \textit{bad} conductors in the normal phase are \textit{good} superconductors and suggest an alternative mechanism of high-temperature pair-of-bogolons--mediated superconductivity.


\textit{In conclusion}, we have studied the transport of electrons coupled with a two-dimensional Bose-condensed dipolar exciton gas via the Coulomb interaction. We calculated the resistivity in the Bloch-Gr\"{u}neisen regime and provided the analytical formulas for the single and two-bogolon scattering channels for GaAs and MoS$_2$ materials and found that two-bogolon scattering is the dominant mechanism in hybrid systems. Furthermore, we suggested an alternative way of electron pairing mediated by a pair of bogolons.


We have been supported by the Institute for Basic Science in Korea (Project No.~IBS-R024-D1) and the Russian Science Foundation (Project No. 17-12-01039).



%
%
%
%


%
%
%
%

\bibliography{library}
\bibliographystyle{apsrev4-1}


%
\begin{widetext}

\appendix



\newpage
\section{SUPPLEMENTAL MATERIAL}
In this Supplemental Material, we derive the low- and high-temperature $T$-dependences of the resistivity of a two-dimensional electron gas (with parabolic dispersion) in a hybrid Bose-Fermi system, extending the Bloch-Gr\"uneisen approach. The single-bogolon scattering is considered in Appendix A while the two-bogolon case is considered in Appendix B. In Appendix C we show the calculationn of the screening effect. Finally, we calculate the phonon contribution in Appendix D.


\section{Appendix A: Bloch-Gr\"uneisen formula for single-bogolon processes}
We start from the Boltzmann equation
\begin{equation}\label{1}
e_0\textbf{E}\cdot\frac{\partial f}{\hbar\partial \textbf{p}}=I\{f\},
\end{equation}
where $\mathbf{p}$ is the wave vector and we will use $p\equiv\abs{\mathbf{p}}$, $\mathbf{E}$ is the perturbing electric field, and $f$ is the distribution function.
The scattering integral is given by
\begin{eqnarray}
\label{2}
I\{f\}&=&-\frac{1}{\hbar}\int\frac{d\textbf{q}d\textbf{p}'}{(2\pi)^2}|M_q|^2\Bigl[N_qf_p(1-f_{p'})\delta(\varepsilon_p-\varepsilon_{p'}+\hbar\omega_q)\delta(\textbf{p}-\textbf{p}'+\textbf{q})\nonumber\\
&{}&{}+(N_q+1)f_p(1-f_{p'})\delta(\varepsilon_p-\varepsilon_{p'}-\hbar\omega_q)\delta(\textbf{p}-\textbf{p}'-\textbf{q})+N_qf_{p'}(1-f_{p})\delta(\varepsilon_{p'}-\varepsilon_{p}+\hbar\omega_q)\delta(\textbf{p}'-\textbf{p}+\textbf{q})\nonumber\\
&{}&{}+(N_q+1)f_{p'}(1-f_p)\delta(\varepsilon_{p'}-\varepsilon_{p}-\hbar\omega_q)\delta(\textbf{p}'-\textbf{p}-\textbf{q})\Bigr].
\end{eqnarray}
(Note, that the sample length $L$ cancels out in the equation above.)

For small enough electric fields, the electron distribution is not substantially different from the equilibrium Fermi distribution, thus it can be presented in the form
\begin{eqnarray}
f=f^0(\varepsilon_p)-\left(-\frac{\partial f^0}{\partial\varepsilon_p}\right)f^{(1)}_\textbf{p},
\end{eqnarray}
where $f^0$ is the equilibrium Fermi-Dirac distribution and $f^{(1)}_\textbf{p}$ has a dimensionality of energy.
Following the steps of the derivation reported in~\cite{Zaitsev:2014aa}, we rewrite:
\begin{eqnarray}
\label{3}
e_0\textbf{E}\cdot\frac{\partial f}{\hbar\partial \textbf{p}}&=&\frac{\hbar e_0}{m}\textbf{E}\cdot\textbf{p}\frac{\partial f^0}{\partial \varepsilon_p}=I\{f^{(1)}_\textbf{p}\},\\\nonumber
I\{f^{(1)}_\textbf{p}\}&=&-\frac{1}{\hbar}\int\frac{d\textbf{q}d\textbf{p}'}{(2\pi)^2}|M_q|^2\frac{1}{\hbar}\frac{\partial N_q}{\partial \omega_q}
\left(f^0(\varepsilon_p)-f^0(\varepsilon_{p'})\right)
\left(f^{(1)}_{\textbf{p}}-f^{(1)}_{\textbf{p}'}\right)
\Bigl[\delta(\varepsilon_{p}-\varepsilon_{p'}-\hbar\omega_q)\delta(\textbf{p}-\textbf{p}'-\textbf{q})\nonumber\\
&{}&{}-\delta(\varepsilon_{p}-\varepsilon_{p'}+\hbar\omega_q)\delta(\textbf{p}-\textbf{p}'+\textbf{q})\Bigr],
\end{eqnarray}
where
$$\frac{\partial N_q}{\partial \omega_q}=-\frac{\hbar}{k_BT}N_q(1+N_q)$$
and $m$ is the effective mass of the electron in 2DEG which has the dispersion $\varepsilon_p=\frac{\hbar^2p^2}{2m}$.

Further we integrate over the electron wave vector $\textbf{p}'$ and find
\begin{eqnarray}
\label{4}
\frac{\hbar e_0}{m}\textbf{E}\cdot\textbf{p}\frac{\partial f^0}{\partial \varepsilon_p}
=&-&\frac{1}{\hbar}\int\frac{d\textbf{q}}{(2\pi)^2}|M_q|^2\frac{1}{\hbar}\frac{\partial N_q}{\partial \omega_q}
\left(f^0(\varepsilon_p)-f^0(\varepsilon_{p}-\hbar\omega_q)\right)
\left(f^{(1)}_{\textbf{p}}-f^{(1)}_{\textbf{p}-\textbf{q}}\right)
\delta(\varepsilon_{\textbf{p}}-\varepsilon_{\textbf{p}-\textbf{q}}-\hbar\omega_\textbf{q})\\
\nonumber
&+&\frac{1}{\hbar}\int\frac{d\textbf{q}}{(2\pi)^2}|M_q|^2\frac{1}{\hbar}\frac{\partial N_q}{\partial \omega_q}
\left(f^0(\varepsilon_p)-f^0(\varepsilon_{p}+\hbar\omega_q)\right)
\left(f^{(1)}_{\textbf{p}}-f^{(1)}_{\textbf{p}+\textbf{q}}\right)
\delta(\varepsilon_{\textbf{p}}-\varepsilon_{\textbf{p}+\textbf{q}}+\hbar\omega_\textbf{q}).
\end{eqnarray}
Let the electric field be directed along the $x$-axis. Then we can use the correction function (which is the correction to the homogeneous distribution) in the form
\begin{eqnarray}
f^{(1)}_\textbf{p}= \frac{\hbar e_0}{m}E_xp_x\tau(\varepsilon_p),
\end{eqnarray}
where $p_F$ is the Fermi wave vector and $\tau(\varepsilon_p)$ is the relaxation time.  
We have
\begin{eqnarray}\label{5}
\hbar p_x\frac{\partial f^0}{\partial \varepsilon_p}
=&-&\frac{1}{\hbar}\int\frac{d\textbf{q}}{(2\pi)^2}|M_q|^2\frac{1}{\hbar}\frac{\partial N_q}{\partial \omega_q}
\left[f^0(\varepsilon_p)-f^0(\varepsilon_{p}-\hbar\omega_q)\right]
\left[\frac{p_x}{k_F}\tau(\varepsilon_p)-\frac{p_x-q_x}{k_F}\tau(\varepsilon_p-\hbar\omega_q)\right]
\delta(\varepsilon_{\textbf{p}}-\varepsilon_{\textbf{p}-\textbf{q}}-\hbar\omega_\textbf{q})\nonumber\\
&+&\frac{1}{\hbar}\int\frac{d\textbf{q}}{(2\pi)^2}|M_q|^2\frac{1}{\hbar}\frac{\partial N_q}{\partial \omega_q}
\left[f^0(\varepsilon_p)-f^0(\varepsilon_{p}+\hbar\omega_q)\right]
\left[\frac{p_x}{k_F}\tau(\varepsilon_p)-\frac{p_x+q_x}{k_F}\tau(\varepsilon_p+\hbar\omega_q)\right]
\delta(\varepsilon_{\textbf{p}}-\varepsilon_{\textbf{p}+\textbf{q}}+\hbar\omega_\textbf{q}).\nonumber
\end{eqnarray}
We now replace the relaxation time by its energy-averaged value~\cite{Ziman:2001aa} $\tau=\tau_0$ and find
\begin{eqnarray}
\label{Boltzmann1}
\hbar p_x\frac{\partial f^0}{\partial \varepsilon_p}
=&-&\tau_0\int \frac{d\textbf{q}}{(2\pi)^2}q_x|M_q|^2\frac{1}{\hbar}\frac{\partial N_q}{\partial \omega_q}
\left[f^0(\varepsilon_p)-f^0(\varepsilon_{p}-\hbar\omega_q)\right]
\delta(\varepsilon_{\textbf{p}}-\varepsilon_{\textbf{p}-\textbf{q}}-\hbar\omega_\textbf{q})\\
\nonumber
&-&\tau_0\int \frac{d\textbf{q}}{(2\pi)^2}q_x|M_q|^2\frac{1}{\hbar}\frac{\partial N_q}{\partial \omega_q}
\left[f^0(\varepsilon_p)-f^0(\varepsilon_{p}+\hbar\omega_q)\right]
\delta(\varepsilon_{\textbf{p}}-\varepsilon_{\textbf{p}+\textbf{q}}+\hbar\omega_\textbf{q}).
\end{eqnarray}

Let us denote the angle between the vectors $\textbf{p}$ and $\textbf{q}$ as $\varphi$ and the angle between the vectors $\textbf{p}$ and $\textbf{E}$ as $\beta$. Then $q_x=q\cos(\varphi+\beta)$ and $p_x=p\cos\beta$. Integrating over $\phi$, we find
\begin{eqnarray}
\label{Eq19}
\int_0^{2\pi} d\phi\cos (\phi+\beta)\delta(\varepsilon_p-\varepsilon_{|\mathbf{p}\pm\mathbf{q}|}\pm\hbar\omega_q)=\frac{2m}{\hbar^2p}\cos\beta\frac{\left(\mp\frac{q}{2p}+\frac{ms}{\hbar p}\right)\Theta\left[1-\left(\mp\frac{q}{2p}+\frac{ms}{\hbar p}\right)^2\right]}{\sqrt{1-\left(\mp\frac{q}{2p}+\frac{ms}{\hbar p}\right)^2}},
\end{eqnarray}
where $\Theta[x]$ is the Heaviside step function. 
To derive Eq.~\eqref{Eq19}, we denoted a new variable $x=\cos\phi$. This implied $d\phi=\mp dx[1-x^2]^{-1/2}$, where the $-$($+$) case is for $0\leq\phi <\pi$ ($\pi\leq\phi<2\pi$).

After integrating over the angle $\phi$, we can integrate Eq.~\eqref{Boltzmann1} over $\xi_p=\varepsilon_p-\mu$, using
\begin{eqnarray}
\int\limits_{-\infty}^{\infty}d\xi_p\frac{\partial f^0}{\partial \varepsilon_p}=-1,\nonumber\\
\int\limits_{-\infty}^{\infty}d\xi_p\left(f^0(\varepsilon_p)-f^0(\varepsilon_{p}\pm\hbar\omega_q)\right)=\pm\hbar\omega_q,
\end{eqnarray}
and putting all electron wave vectors to be $p=k_F$.

The resistivity is inversely proportional to the scattering time,
\begin{eqnarray}
\label{rho1}
\rho\propto\frac{1}{\tau_0}&=&\frac{m\xi_I^2}{\hbar k_F^3M}\frac{1}{k_BT}\int_0^\infty \frac{dq}{(2\pi)^2} \frac{q^3e^{-2ql}}{\epsilon(q)^2}(\Gamma_+-\Gamma_-)_{k_F}N_q(1+N_q),
\end{eqnarray}
where we introduced $\xi_I= e_0^2d\sqrt{n_c}/2\epsilon$ and
\begin{eqnarray}
\label{gamma}
\Gamma_{\pm}=\frac{\left(\mp\frac{q}{2p}+\frac{ms}{\hbar p}\right)\Theta\left[1-\left(\mp\frac{q}{2p}+\frac{ms}{\hbar p}\right)^2\right]}{\sqrt{1-\left(\mp\frac{q}{2p}+\frac{ms}{\hbar p}\right)^2}}.
\end{eqnarray}
Here, $\epsilon(q)$ is the static screening given by
\begin{eqnarray}
\epsilon(q)=\left(1+\frac{2}{a_Bq}\right)\left(1+\frac{1}{q^2\xi^2}\right),
\end{eqnarray}
where $a_B$ is the Bohr radius and, recall, $\xi$ is the healing length of the condensate.

The subscript $k_F$ in the expression $(\Gamma_--\Gamma_+)_{k_F}$ in Eq.~\eqref{rho1} means that all the electron wave vectors $p$ are to be substituted by the Fermi value $k_F$.
%
%
%
%
%

We now introduce a new dimensionless variable
\begin{eqnarray}
\label{u}
u=\frac{\hbar sq}{k_BT}
\end{eqnarray}
in Eq.~\eqref{rho1} and obtain
\begin{eqnarray}
\label{rho2}
\frac{1}{\tau_0}&=&\frac{m\xi_I^2}{\hbar k_F^3M}\frac{(k_BT)^3}{(\hbar s)^4}\int_0^\infty \frac{du}{(2\pi)^2} \frac{u^3e^{(1-2\tilde{l})u}}{(e^u-1)^2}(\Gamma_+-\Gamma_-)_{k_F},
\end{eqnarray}
where
\begin{eqnarray}
\Tilde{l}=\frac{lk_BT}{\hbar s}\sim\frac{k_BT}{10\mbox{ meV}}
\end{eqnarray}
and we used $s=10^5$ m/s and $l= 5.0\times 10^{-8}$ m/s. 
Note that the room temperature is $k_BT_R\sim 26$ meV, so that for temperatures far less than the room temperature we have $\Tilde{l}\ll 1$. Hence we can replace
\begin{eqnarray}
\label{approx1}
e^{(1-2\tilde{l})u}\rightarrow e^u.
\end{eqnarray}
To keep things general, we instead expand
\begin{eqnarray}
\label{taylor}
e^{-2\tilde{l}u}=\sum_{n=0}^\infty\frac{(-1)^n(2\tilde{l}u)^n}{n!}.
\end{eqnarray}

Let us now look at the argument of the Heaviside theta function in Eq.~\eqref{gamma}. The roots for both the cases are
\begin{eqnarray}
q=\pm 2k_F-\frac{2ms}{\hbar}\approx\pm2k_F,
\end{eqnarray}
which means that the Heaviside theta function is non-zero in the integration range
\begin{eqnarray}
0\leq q\lesssim 2k_F,
\end{eqnarray}
or in terms of $u$ [introduced in Eq.~\eqref{u}],
\begin{eqnarray}
\label{lambda}
0\leq u<\frac{T_\textrm{BG}}{T}\equiv\Lambda,
\end{eqnarray}
where $T_\textrm{BG}=2\hbar sk_F/k_B$ is the Bloch-Gr\"{u}neisen temperature for bogolons.

%

For large $u$ (or $q$), the factors $\Gamma_\pm$ in Eq.~\eqref{rho2} approach constant values. In the mean time, the term $u^4\exp({-u})$ rapidly goes to zero for $u>10$.
Therefore we can remove the theta function in Eq.~\eqref{gamma} and this incurs only a small (imaginary) error. The term inside the square root of Eq.~\eqref{gamma} can be rewritten as
\begin{eqnarray}
\label{sqrt}
1-\left(\mp\frac{q}{2p}+\frac{ms}{\hbar p}\right)^2&\approx &\frac{1}{4k_F^2}(2k_F-q)(2k_F+q)\nonumber\\
&=&\left(\frac{k_BT}{2k_F\hbar s}\right)^2(\Lambda-u)(\Lambda+u),
\end{eqnarray}
where $\Lambda$ does depend on $T$, as was defined in Eq.~\eqref{lambda}. However, for $T\ll T_\textrm{BG}$, due to the factor $\exp({-u})$ we can simply replace $\Lambda\sim 10$ (or greater) without significantly affecting the result. 

At low temperatures $T\rightarrow 0$, the screening factor reads
\begin{eqnarray}
\epsilon(u)=1+\frac{\hbar^2s^2}{k_B^2T^2\xi^2}\frac{1}{u^2}+\frac{2\hbar s}{k_BTu}+\frac{2(\hbar s)^3}{(k_BT)^3\xi^2}\frac{1}{u^3}\approx\frac{2(\hbar s)^3}{(k_BT)^3\xi^2}\frac{1}{u^3},
\end{eqnarray}
where we used Eq.~\eqref{u} to trade $q$ with $u$.

Using Eqs.~\eqref{taylor}, ~\eqref{sqrt}, and the arguments presented above, which allow us to remove the Theta function, we find
\begin{eqnarray}
\rho=\frac{\pi\hbar^2}{e_0^2E_F}\frac{1}{\tau_0}=\frac{\pi\hbar^3\xi_I^2\xi^6}{4e_0^2ME_F}\sum_{n=0}^\infty\frac{(-2)^nl^n\gamma_n}{n!(\hbar s)^{n+10}}(k_BT)^{n+9},
\end{eqnarray}
where
\begin{eqnarray}
\label{EqMath}
\gamma_n=\int_0^\Lambda\frac{du}{(2\pi)^2}\frac{e^uu^{n+9}}{(e^u-1)^2\sqrt{(\Lambda-u)(\Lambda+u)}}.
\end{eqnarray}
This dimensionless integral can be evaluated in closed form when we note that (i) $\Lambda\gg 1$ and (ii) that, due to the exponential factors in the integrand, the relevant contribution to the integral comes from $0<u\lesssim 1$, so that
\begin{eqnarray}
\label{EqMath2}
\gamma_n\approx\frac{1}{\Lambda}\int_0^\infty\frac{du}{(2\pi)^2}\frac{e^uu^{n+9}}{(e^u-1)^2}=\frac{(n+9)!}{(2\pi)^2}\zeta(n+9)\frac{T}{T_\textrm{BG}}.
\end{eqnarray}
The leading term at low temperatures are then given by
\begin{eqnarray}
\rho\approx\frac{9!\pi\hbar^3\xi_I^2\xi^6}{4(2\pi)^2e_0^2ME_Fk_BT_\textrm{BG}}\zeta(9)\left(\frac{k_BT}{\hbar s}\right)^{10}.
\end{eqnarray}
Hence, at low temperatures the resistivity behaves as $\rho\propto T^{10}$ (with screening).

\section{Appendix B: Bloch-Gr\"uneisen formula for two-bogolon processes}

The starting equation is
\begin{eqnarray}
\label{EqKin}
e\mathbf{E}\cdot\frac{df_p}{\hbar d\mathbf{p}}=I\{f_p\}.
\end{eqnarray}
We consider the Hamiltonian
\begin{equation}\label{1}
H=\sum_{\mathbf{k},\mathbf{p}',\mathbf{q},\mathbf{q}'}g(\textbf{k})c^+_{\textbf{p}'}c_{\textbf{p}}\varphi^+_{\textbf{q}'}\varphi_{\textbf{q}}
\delta(\textbf{p}'-\textbf{p}-\textbf{k})\delta(\textbf{q}'-\textbf{q}+\textbf{k}),
\end{equation}
where
\begin{gather}\label{2}
\varphi^+_{\textbf{q}'}\varphi_{\textbf{q}}=(u_{\textbf{q}'}b^+_{\textbf{q}'}+v_{\textbf{q}'}b_{-\textbf{q}'})
(u_{\textbf{q}}b_{\textbf{q}}+v_{\textbf{q}}b^+_{-\textbf{q}})
=u_{\textbf{q}'}b^+_{\textbf{q}'}u_{\textbf{q}}b_{\textbf{q}}+u_{\textbf{q}'}b^+_{\textbf{q}'}v_{\textbf{q}}b^+_{-\textbf{q}}+
v_{\textbf{q}'}b_{-\textbf{q}'}u_{\textbf{q}}b_{\textbf{q}}+v_{\textbf{q}'}b_{-\textbf{q}'}v_{\textbf{q}}b^+_{-\textbf{q}},
\end{gather}
as in our manuscript.
Thus, the collision integral reads
\begin{gather}\label{3}
I\{f_p\}=I_1-I_2,~\textrm{where}\\
\nonumber
I_1=-\sum_{\mathbf{k},\mathbf{p}',\mathbf{q},\mathbf{q}'}|g(\textbf{k})|^2f_p(1-f_{p'})\delta(\textbf{p}'-\textbf{p}-\textbf{k})\delta(\textbf{q}'-\textbf{q}+\textbf{k})\times\\\nonumber
\times\Bigl[u^2_{\textbf{q}'}u^2_{\textbf{q}}(N_{\textbf{q}'}+1)N_{\textbf{q}}
\delta(\varepsilon_{\textbf{p}'}-\varepsilon_{\textbf{p}}+\omega_{\textbf{q}'}-\omega_{\textbf{q}})
+u^2_{\textbf{q}'}v^2_{\textbf{q}}(N_{\textbf{q}'}+1)(N_{-\textbf{q}}+1)
\delta(\varepsilon_{\textbf{p}'}-\varepsilon_{\textbf{p}}+\omega_{\textbf{q}'}+\omega_{-\textbf{q}})+\\\nonumber
+v^2_{\textbf{q}'}u^2_{\textbf{q}}N_{-\textbf{q}'}N_{\textbf{q}}
\delta(\varepsilon_{\textbf{p}'}-\varepsilon_{\textbf{p}}-\omega_{-\textbf{q}'}-\omega_{\textbf{q}})
+v^2_{\textbf{q}'}v^2_{\textbf{q}}N_{-\textbf{q}'}(N_{-\textbf{q}}+1)
\delta(\varepsilon_{\textbf{p}'}-\varepsilon_{\textbf{p}}-\omega_{-\textbf{q}'}+\omega_{-\textbf{q}})\Bigr],\\
\nonumber
I_2=-\sum_{\mathbf{k},\mathbf{p}',\mathbf{q},\mathbf{q}'}|g(\textbf{k})|^2f_{p'}(1-f_{p})\delta(\textbf{p}-\textbf{p}'-\textbf{k})\delta(\textbf{q}'-\textbf{q}+\textbf{k})\times\\\nonumber
\times\Bigl[u^2_{\textbf{q}'}u^2_{\textbf{q}}(N_{\textbf{q}'}+1)N_{\textbf{q}}
\delta(\varepsilon_{\textbf{p}}-\varepsilon_{\textbf{p}'}+\omega_{\textbf{q}'}-\omega_{\textbf{q}})
+u^2_{\textbf{q}'}v^2_{\textbf{q}}(N_{\textbf{q}'}+1)(N_{-\textbf{q}}+1)
\delta(\varepsilon_{\textbf{p}}-\varepsilon_{\textbf{p}'}+\omega_{\textbf{q}'}+\omega_{-\textbf{q}})+\\\nonumber
+v^2_{\textbf{q}'}u^2_{\textbf{q}}N_{-\textbf{q}'}N_{\textbf{q}}
\delta(\varepsilon_{\textbf{p}}-\varepsilon_{\textbf{p}'}-\omega_{-\textbf{q}'}-\omega_{\textbf{q}})
+v^2_{\textbf{q}'}v^2_{\textbf{q}}N_{-\textbf{q}'}(N_{-\textbf{q}}+1)
\delta(\varepsilon_{\textbf{p}}-\varepsilon_{\textbf{p}'}-\omega_{-\textbf{q}'}+\omega_{-\textbf{q}})\Bigr],
\end{gather}
where we assume that $N_\mathbf{x}$ are equilibrium Bose distribution functions. In $I_2$ we can change the signs of the vectors: $\textbf{k}\rightarrow-\textbf{k},\,\textbf{q}\rightarrow-\textbf{q},\,\textbf{q}'\rightarrow-\textbf{q}'$. Taking into account that the distribution functions and energies only depend on the absolute value of the wave vectors, we find
\begin{gather}\label{4}
I_1=-\sum_{\mathbf{k},\mathbf{p}',\mathbf{q},\mathbf{q}'}|g(\textbf{k})|^2f_p(1-f_{p'})\delta(\textbf{p}'-\textbf{p}-\textbf{k})\delta(\textbf{q}'-\textbf{q}+\textbf{k})\times\\\nonumber
\times\Bigl[u^2_{\textbf{q}'}u^2_{\textbf{q}}(N_{\textbf{q}'}+1)N_{\textbf{q}}
\delta(\varepsilon_{\textbf{p}'}-\varepsilon_{\textbf{p}}+\omega_{\textbf{q}'}-\omega_{\textbf{q}})
+u^2_{\textbf{q}'}v^2_{\textbf{q}}(N_{\textbf{q}'}+1)(N_{\textbf{q}}+1)
\delta(\varepsilon_{\textbf{p}'}-\varepsilon_{\textbf{p}}+\omega_{\textbf{q}'}+\omega_{\textbf{q}})+\\\nonumber
+v^2_{\textbf{q}'}u^2_{\textbf{q}}N_{\textbf{q}'}N_{\textbf{q}}
\delta(\varepsilon_{\textbf{p}'}-\varepsilon_{\textbf{p}}-\omega_{\textbf{q}'}-\omega_{\textbf{q}})
+v^2_{\textbf{q}'}v^2_{\textbf{q}}N_{\textbf{q}'}(N_{\textbf{q}}+1)
\delta(\varepsilon_{\textbf{p}'}-\varepsilon_{\textbf{p}}-\omega_{\textbf{q}'}+\omega_{\textbf{q}})\Bigr],\\
\nonumber
I_2=-\sum_{\mathbf{k},\mathbf{p}',\mathbf{q},\mathbf{q}'}|g(\textbf{k})|^2f_{p'}(1-f_{p})\delta(-\textbf{p}+\textbf{p}'+\textbf{k})\delta(-\textbf{q}'+\textbf{q}-\textbf{k})\times\\\nonumber
\times\Bigl[u^2_{\textbf{q}'}u^2_{\textbf{q}}(N_{\textbf{q}'}+1)N_{\textbf{q}}
\delta(\varepsilon_{\textbf{p}}-\varepsilon_{\textbf{p}'}+\omega_{\textbf{q}'}-\omega_{\textbf{q}})
+u^2_{\textbf{q}'}v^2_{\textbf{q}}(N_{\textbf{q}'}+1)(N_{\textbf{q}}+1)
\delta(\varepsilon_{\textbf{p}}-\varepsilon_{\textbf{p}'}+\omega_{\textbf{q}'}+\omega_{\textbf{q}})+\\\nonumber
+v^2_{\textbf{q}'}u^2_{\textbf{q}}N_{\textbf{q}'}N_{\textbf{q}}
\delta(\varepsilon_{\textbf{p}}-\varepsilon_{\textbf{p}'}-\omega_{\textbf{q}'}-\omega_{\textbf{q}})
+v^2_{\textbf{q}'}v^2_{\textbf{q}}N_{\textbf{q}'}(N_{\textbf{q}}+1)
\delta(\varepsilon_{\textbf{p}}-\varepsilon_{\textbf{p}'}-\omega_{\textbf{q}'}+\omega_{\textbf{q}})\Bigr].
\end{gather}
We see that the delta-functions describing the momentum conservation are the same. We also use that for linear spectrum of bogolons,  $u^2_{\textbf{q}'}u^2_{\textbf{q}}=u^2_{\textbf{q}'}v^2_{\textbf{q}}=v^2_{\textbf{q}'}v^2_{\textbf{q}}=v^2_{\textbf{q}'}v^2_{\textbf{q}}$. It yields:
\begin{eqnarray}\label{5}
I_1-I_2=-\sum_{\mathbf{k},\mathbf{p}',\mathbf{q},\mathbf{q}'}u^2_{\textbf{q}'}u^2_{\textbf{q}}|g(\textbf{k})|^2\delta(\textbf{p}'-\textbf{p}-\textbf{k})\delta(\textbf{q}'-\textbf{q}+\textbf{k})\\
\nonumber
\times
\left\{
\Bigr[N_{\textbf{q}'}N_{\textbf{q}}f_p(1-f_{p'})-(N_{\textbf{q}'}+1)(N_{\textbf{q}}+1)f_{p'}(1-f_{p})\Bigl]
\delta(\varepsilon_{\textbf{p}'}-\varepsilon_{\textbf{p}}-\omega_{\textbf{q}'}-\omega_{\textbf{q}})+\right.\\
\nonumber
\left.+\Bigl[(N_{\textbf{q}'}+1)(N_{\textbf{q}}+1)f_p(1-f_{p'})-N_{\textbf{q}'}N_{\textbf{q}}f_{p'}(1-f_{p})\Bigr]
\delta(\varepsilon_{\textbf{p}'}-\varepsilon_{\textbf{p}}+\omega_{\textbf{q}'}+\omega_{\textbf{q}})+\right.\\
\nonumber
\left.+\Bigr[N_{\textbf{q}'}(N_{\textbf{q}}+1)f_p(1-f_{p'})-(N_{\textbf{q}'}+1)N_{\textbf{q}}f_{p'}(1-f_{p})\Bigl]
\delta(\varepsilon_{\textbf{p}'}-\varepsilon_{\textbf{p}}-\omega_{\textbf{q}'}+\omega_{\textbf{q}})+\right.\\
\nonumber
\left.+\Bigl[(N_{\textbf{q}'}+1)N_{\textbf{q}}f_p(1-f_{p'})-N_{\textbf{q}'}(N_{\textbf{q}}+1)f_{p'}(1-f_{p})\Bigr]
\delta(\varepsilon_{\textbf{p}'}-\varepsilon_{\textbf{p}}+\omega_{\textbf{q}'}-\omega_{\textbf{q}})
\right\}.
\end{eqnarray}

The sums in~\eqref{5} can be replaced by integrals in the continuous limit. Then we can consider the variations of such integrals over $f_p$ and $f_{p^\prime}$~\cite{Zaitsev:2014aa}. For example, let us consider the terms in the square brackets in the second line of Eq.~\eqref{5}: $N_{\textbf{q}'}N_{\textbf{q}}f_p(1-f_{p'})-(N_{\textbf{q}'}+1)(N_{\textbf{q}}+1)f_{p'}(1-f_{p})$. Taking a variation over $f_p$ we find:
\begin{eqnarray}
\label{EqBol}
\delta f_p\left\{(1-f_{p^\prime})N_qN_{q^\prime}+f_{p^\prime}(N_q+1)(N_{q^\prime}+1)\right\}
=
\delta f_pN_qN_{q^\prime}n_F(p^\prime)\left\{\exp\left(\frac{\xi_{p^{\prime}}}{k_BT}\right)+\exp\left(\frac{\hbar s (q+q^\prime)}{k_BT}\right)\right\}\\
\nonumber
=\delta f_pN_qN_{q^\prime}n_F(p^\prime)
\exp\left(\frac{\hbar s (q+q^\prime)}{k_BT}\right)
\left\{\exp\left(\frac{\xi_{p^{\prime}}-\hbar s (q+q^\prime)}{k_BT}\right)+1\right\}
=
\delta f_p(N_q+1)(N_{q^\prime}+1)n_F(p^\prime)\frac{1}{n_F(p)},
\end{eqnarray}
where we denoted the equilibrium Fermi distribution as $n_F(p)\equiv n_F(\xi_p)$ and $\xi_p=\varepsilon_p-\mu$, where $\mu$ is the chemical potential. In the last equality we also used the energy conservation: $\xi_{p}=\xi_{p^{\prime}}-\hbar s (q+q^\prime)$ (legitimate for this particular term).

Further we assume that the distribution function of electrons $f_p$ is close to the equilibrium one, thus expanding $f_p=n_F(\xi_p)+\delta f_p=n_F(\xi_p)+(\partial n_F(p)/\partial \xi_p)\varphi_p=n_F(\xi_p)+\frac{1}{k_BT}n_F(\xi_p)[1-n_F(\xi_p)]\varphi_p$, where we introduce the correction $\varphi_p$. Then~\eqref{EqBol} (after some algebra) turns into
\begin{eqnarray}
\label{EqPhi}
\frac{\varphi_p}{k_BT}n_F(p)(1-n_F(p))
(N_q+1)(N_{q^\prime}+1)n_F(p^\prime)\frac{1}{n_F(p)}
=\frac{\varphi_p}{k_BT}(n_F(p)-n_F(p^\prime))
(N_q+1)(N_{q^\prime}+1)N_{q+q^\prime}.
\end{eqnarray}

In similar fashion we treat the variation of the first line in Eq.~\eqref{5}  over $f_{p^\prime}$ and find:
\begin{eqnarray}
\label{EqPhiPrime}
-\frac{\varphi_{p^\prime}}{k_BT}(n_F(p)-n_F(p^\prime))
N_qN_{q^\prime}(N_{q+q^\prime}+1).
\end{eqnarray}
Discovering that $N_qN_{q^\prime}(N_{q+q^\prime}+1)=(N_q+1)(N_{q^\prime}+1)N_{q+q^\prime}$, which means that $\varphi_p$ and $\varphi_{p^\prime}$ in~\eqref{EqPhi} and~\eqref{EqPhiPrime} have equivalent prefactors, we find the first (out of four) term to enter our target expression:
\begin{eqnarray}
-\frac{\varphi_p-\varphi_{p^\prime}}{k_BT}(n_F(p)-n_F(p^\prime))
N_qN_{q^\prime}(N_{q+q^\prime}+1)\delta(\varepsilon_{\textbf{p}'}-\varepsilon_{\textbf{p}}-\omega_{\textbf{q}'}-\omega_{\textbf{q}}).
\end{eqnarray}

Repeating a similar variation procedure with all the other terms in Eq.~\eqref{5}, we find the total formula:
\begin{gather}\label{EqZaiatz}
I_1-I_2=-\sum_{\mathbf{k},\mathbf{p}',\mathbf{q},\mathbf{q}'}u^2_{\textbf{q}'}u^2_{\textbf{q}}|g(\textbf{k})|^2
\frac{\varphi_{\mathbf{p}}-\varphi_{\mathbf{p^\prime}}}{k_BT}
[n_F(\mathbf{p})-n_F(\mathbf{p}^\prime)]
\delta(\textbf{p}'-\textbf{p}-\textbf{k})\delta(\textbf{q}'-\textbf{q}+\textbf{k})\\
\nonumber
\times
\{
N_{q}N_{q^\prime}(N_{q+q^\prime}+1)
\Bigr[\delta(\varepsilon_{\textbf{p}'}-\varepsilon_{\textbf{p}}-\omega_{\textbf{q}'}-\omega_{\textbf{q}})
-
\delta(\varepsilon_{\textbf{p}'}-\varepsilon_{\textbf{p}}+\omega_{\textbf{q}'}+\omega_{\textbf{q}})\Bigl]+\\
\nonumber
+
(N_{q}+1)N_{q^\prime}(N_{q^\prime-q}+1)
\Bigr[\delta(\varepsilon_{\textbf{p}'}-\varepsilon_{\textbf{p}}-\omega_{\textbf{q}'}+\omega_{\textbf{q}})
-
\delta(\varepsilon_{\textbf{p}'}-\varepsilon_{\textbf{p}}+\omega_{\textbf{q}'}-\omega_{\textbf{q}})\Bigl]
\}.
\end{gather}

This expression can be presented in the form
\begin{gather}\label{6}
I_1-I_2=-\sum_{\mathbf{k},\mathbf{p}'}|g(\textbf{k})|^2\frac{\varphi_{\mathbf{p}}-\varphi_{\mathbf{p^\prime}}}{k_BT}
[n_F(\mathbf{p})-n_F(\mathbf{p}^\prime)]
\delta(\textbf{p}'-\textbf{p}-\textbf{k})\int d \epsilon \delta(\varepsilon_{\textbf{p}'}-\varepsilon_{\textbf{p}}-\epsilon)F(\textbf{k},\epsilon),\\\nonumber
F(\textbf{k},\epsilon)=\sum_{\mathbf{q},\mathbf{q}'} u^2_{\textbf{q}'}u^2_{\textbf{q}}
\delta(\textbf{q}'-\textbf{q}+\textbf{k})
\Bigl(
N_{q}N_{q^\prime}(N_{q+q^\prime}+1)
\Bigl[\delta(\epsilon-\omega_{\textbf{q}'}-\omega_{\textbf{q}})
-
\delta(\epsilon+\omega_{\textbf{q}'}+\omega_{\textbf{q}})\Bigl]+\\
\nonumber
+
(N_{q}+1)N_{q^\prime}(N_{q^\prime-q}+1)
\Bigr[\delta(\epsilon-\omega_{\textbf{q}'}+\omega_{\textbf{q}})
-
\delta(\epsilon+\omega_{\textbf{q}'}-\omega_{\textbf{q}})\Bigr]
\Bigr).
\end{gather}
Now we can integrate over $\textbf{p}',\textbf{q}'$. Using the momentum-conserving delta functions, we find:
\begin{gather}\label{7}
I=-\sum_{\mathbf{k}}\int d \epsilon |g(\textbf{k})|^2\frac{\varphi_{\mathbf{p}}-\varphi_{\textbf{p}+\textbf{k}}}{T}
[n_F(\varepsilon_{\textbf{p}})-n_F(\varepsilon_{\textbf{p}}+\epsilon)]
 \delta(\varepsilon_{\textbf{p}+\textbf{k}}-\varepsilon_{\textbf{p}}-\epsilon)F(\textbf{k},\epsilon);\\\nonumber
F(\textbf{k},\epsilon)=\sum_{\mathbf{q}} u^2_{|\textbf{q}-\textbf{k}|}u^2_{\textbf{q}}
\Bigl(
N_{q}N_{|\textbf{q}-\textbf{k}|}(N_{q+|\textbf{q}-\textbf{k}|}+1)
\Bigl[\delta(\epsilon-\omega_{|\textbf{q}-\textbf{k}|}-\omega_{\textbf{q}})
-
\delta(\epsilon+\omega_{|\textbf{q}-\textbf{k}|}+\omega_{\textbf{q}})\Bigl]+\\
\nonumber
+
(N_{q}+1)N_{|\textbf{q}-\textbf{k}|}(N_{|\textbf{q}-\textbf{k}|-q}+1)
\Bigr[\delta(\epsilon-\omega_{|\textbf{q}-\textbf{k}|}+\omega_{\textbf{q}})
-
\delta(\epsilon+\omega_{|\textbf{q}-\textbf{k}|}-\omega_{\textbf{q}})\Bigr]
\Bigr).
\end{gather}
Let us consider the function $F(\mathbf{k},\epsilon)$. We make a replacement $\mathbf{q}\rightarrow \mathbf{q}+\mathbf{k}$ to find:
\begin{gather}
F(\textbf{k},\epsilon)=\sum_{\mathbf{q}} u^2_{q}u^2_{|\textbf{q}+\textbf{k}|}
\Bigl\{
N_{|\textbf{q}+\textbf{k}|}N_{q}(N_{|\textbf{q}+\textbf{k}|+q}+1)
\Bigl[\delta(\epsilon-\omega_{q}-\omega_{|\textbf{q}+\textbf{k}|})
-
\delta(\epsilon+\omega_{q}+\omega_{|\textbf{q}+\textbf{k}|})\Bigl]+\\
\nonumber
+
(N_{|\textbf{q}+\textbf{k}|}+1)N_{q}(N_{q-|\textbf{q}+\textbf{k}|}+1)
\Bigr[\delta(\epsilon-\omega_{q}+\omega_{|\textbf{q}+\textbf{k}|})
-
\delta(\epsilon+\omega_{q}-\omega_{|\textbf{q}+\textbf{k}|})\Bigr]
\Bigr\}.
\end{gather}
Furthermore we switch from summation to integration $\sum_{\mathbf{q}}\rightarrow\int d\mathbf{q}$, and we introduce a variable $q_1=|\textbf{q}+\textbf{k}|$. Then in $\int d\mathbf{q}$ we will integrate over $q$ and $q_1$ instead of $q$ and the angle between the vectors, using:
\begin{eqnarray}
\int\frac{d\mathbf{q}}{2\pi}=\frac{4}{(2\pi)^2}\int_0^\infty qdq\int_{|q-k|}^{q+k}q_1dq_1
\frac{1}{\sqrt{[(q+k)^2-q_1^2][q_1^2-(q-k)^2]}}.
\end{eqnarray}

It gives
\begin{gather}
F(\textbf{k},\epsilon)=\frac{4}{(2\pi)^2}\int_0^\infty 
qdq 
u^2_{q}
\int_{|q-k|}^{q+k}
q_1dq_1
u^2_{q_1}
\frac{1}{\sqrt{[(q+k)^2-q_1^2][q_1^2-(q-k)^2]}}\\
\nonumber
\times\Bigl\{
N_{q_1}N_{q}(N_{q_1+q}+1)
\Bigl[\delta(\epsilon-\omega_{q}-\omega_{q_1})
-
\delta(\epsilon+\omega_{q}+\omega_{q_1})\Bigl]+\\
\nonumber
+
(N_{q_1}+1)N_{q}(N_{q-q_1}+1)
\Bigr[\delta(\epsilon-\omega_{q}+\omega_{q_1})
-
\delta(\epsilon+\omega_{q}-\omega_{q_1})\Bigr]
\Bigr\}.
\end{gather}
Now we can use the definitions of $u_q$~\cite{Giorgini:1998aa} and linear bogolon dispersions to find:
\begin{gather}
F(\textbf{k},\epsilon)=\frac{4}{(2\pi)^2}\frac{(ms)^2}{4}
\int_0^\infty dq 
\int_{|q-k|}^{q+k} dq_1
\frac{1}{\sqrt{[(q+k)^2-q_1^2][q_1^2-(q-k)^2]}}\\
\nonumber
\times
\Bigl\{
N_{q_1}N_{q}(N_{q_1+q}+1)
\Bigl[\delta(\epsilon-s(q+q_1))
-
\delta(\epsilon+s(q+q_1))\Bigl]+\\
\nonumber
+
(N_{q_1}+1)N_{q}(N_{q-q_1}+1)
\Bigr[\delta(\epsilon-s(q-q_1))
-
\delta(\epsilon+s(q-q_1))\Bigr]
\Bigr\}.
\end{gather}
For convenience we denote new variables $x=s(q+q_1)$ and $y=-s(q-q_1)$ and we introduce the cut-off $sL^{-1}$ in the integrals. This infrared cut-off is necessary for the convergence of the final integral as can be seen later on. We emphasize that this cut-off has a physical grounding: It  means that the momentum integration can not include fluctuations with wavelengthes larger than the sample size $L$. It yields
\begin{gather}
F(\textbf{k},\epsilon)=\frac{1}{2}\left(\frac{ms}{2\pi}\right)^2
\int_{sk+sL^{-1}}^\infty \frac{dx}{\sqrt{x^2-s^2k^2}}
\int_{-sk+sL^{-1}}^{sk-sL^{-1}} \frac{dy}{\sqrt{s^2k^2-y^2}}
\\
\nonumber
\times
\Bigl\{
N\left(\frac{x+y}{2s}\right)N\left(\frac{x-y}{2s}\right)(N\left(\frac{x}{s}\right)+1)
\Bigl[\delta(\epsilon-x)
-
\delta(\epsilon+x)\Bigl]+\\
\nonumber
+
(N\left(\frac{x+y}{2s}\right)+1)N\left(\frac{x-y}{2s}\right)(N\left(\frac{-y}{s}\right)+1)
\Bigr[\delta(\epsilon+y)
-
\delta(\epsilon-y)\Bigr]
\Bigr\}.
\end{gather}
We exchange $y\rightarrow-y$ to find:
\begin{gather}
F(\textbf{k},\epsilon)=\frac{1}{2}\left(\frac{ms}{2\pi}\right)^2
\int_{sk+sL^{-1}}^\infty \frac{dx}{\sqrt{x^2-s^2k^2}}
\int_{-sk+sL^{-1}}^{sk-sL^{-1}} \frac{dy}{\sqrt{s^2k^2-y^2}}
\\
\nonumber
\times
\Bigl\{
N\left(\frac{x+y}{2s}\right)N\left(\frac{x-y}{2s}\right)(N\left(\frac{x}{s}\right)+1)
\Bigl[\delta(\epsilon-x)
-
\delta(\epsilon+x)\Bigl]+\\
\nonumber
+
(N\left(\frac{x-y}{2s}\right)+1)N\left(\frac{x+y}{2s}\right)(N\left(\frac{y}{s}\right)+1)
\Bigr[\delta(\epsilon-y)
-
\delta(\epsilon+y)\Bigr]
\Bigr\}.
\end{gather}
Now we can split the function $F(\textbf{k},\epsilon)$ into two functions $F_1(\textbf{k},\epsilon)$ and $F_2(\textbf{k},\epsilon)$ the following way (thus $F(\textbf{k},\epsilon)=F_1(\textbf{k},\epsilon)+F_2(\textbf{k},\epsilon)$):
\begin{gather}
F_1(\textbf{k},\epsilon)=\frac{1}{2}\left(\frac{ms}{2\pi}\right)^2
\int_{sk+sL^{-1}}^\infty \frac{dx}{\sqrt{x^2-s^2k^2}}
\int_{-sk+sL^{-1}}^{sk-sL^{-1}} \frac{dy}{\sqrt{s^2k^2-y^2}}
\\
\nonumber
\times\Bigl(
N\left(\frac{x+y}{2s}\right)N\left(\frac{x-y}{2s}\right)(N\left(\frac{x}{s}\right)+1)
\Bigl[\delta(\epsilon-x)
-
\delta(\epsilon+x)\Bigl];
\\
\nonumber
F_2(\textbf{k},\epsilon)=\frac{1}{2}\left(\frac{ms}{2\pi}\right)^2
\int_{sk+sL^{-1}}^\infty \frac{dx}{\sqrt{x^2-s^2k^2}}
\int_{-sk+sL^{-1}}^{sk-sL^{-1}} \frac{dy}{\sqrt{s^2k^2-y^2}}
\\
\nonumber
\times\Bigl(N\left(\frac{x-y}{2s}\right)+1)N\left(\frac{x+y}{2s}\right)(N\left(\frac{y}{s}\right)+1)
\Bigr[\delta(\epsilon-y)
-
\delta(\epsilon+y)\Bigr]
\Bigr),
\end{gather}
thus we can separately perform the $x$- and $y$-integrations using the delta-functions.

Let us start consideration with $F_1(\textbf{k},\epsilon)$. Since the integration is performed over $x>0$, we can use the relation $\delta(\epsilon-x)-\delta(\epsilon+x)=\textrm{sgn}(\epsilon)\delta(x-|\epsilon|)$ to find: 
\begin{gather}
F_1(\textbf{k},\epsilon)=\mathrm{sgn}(\epsilon)\frac{1}{2}\left(\frac{ms}{2\pi}\right)^2\frac{\Theta[|\epsilon|-sk]}{\sqrt{\epsilon^2-s^2k^2}}
\int_{-+sL^{-1}}^{sk-sL^{-1}} \frac{dy}{\sqrt{s^2k^2-y^2}}
\\
\nonumber
\times
N\left(\frac{|\epsilon|+y}{2s}\right)N\left(\frac{|\epsilon|-y}{2s}\right)(N\left(\frac{|\epsilon|}{s}\right)+1),
\end{gather}
and using another variable $y=skz$, we find:
\begin{gather}
F_1(\textbf{k},\epsilon)=\frac{\mathrm{sgn}(\epsilon)}{2}\left(\frac{ms}{2\pi}\right)^2\frac{e^{\frac{|\epsilon|}{2T}}}{e^{\frac{|\epsilon|}{T}}-1}\frac{\Theta[|\epsilon|-sk]}{\sqrt{\epsilon^2-s^2k^2}}
\int_{0}^{1-L^{-1}/k} \frac{dz}{\sqrt{1-z^2}}
\frac{1}{\mathrm{cosh}
\left(\frac{|\epsilon|}{2T}\right)
-
\mathrm{cosh}\left(\frac{sk}{2T}z\right)}.
\end{gather}

Now let us take care of the function $F_2(\textbf{k},\epsilon)$. The integral $\int_{-sk}^{sk}$ we can split on two: $\int_{-sk}^{0}$ and $\int_{0}^{sk}$. In the first one we do a replacement $y\rightarrow-y$. After, we combine the two terms and find:
\begin{eqnarray}
F_2(\textbf{k},\epsilon)=\frac{1}{2}\left(\frac{ms}{2\pi}\right)^2
\int_{sk+sL^{-1}}^\infty \frac{dx}{\sqrt{x^2-s^2k^2}}
\int_{0}^{sk-sL^{-1}} \frac{dy}{\sqrt{s^2k^2-y^2}}
\Bigr[\delta(\epsilon-y)
-
\delta(\epsilon+y)\Bigr]
\\
\nonumber
\times\left\{(N\left(\frac{x-y}{2s}\right)+1)N\left(\frac{x+y}{2s}\right)(N\left(\frac{y}{s}\right)+1)
-
(N\left(\frac{x+y}{2s}\right)+1)N\left(\frac{x-y}{2s}\right)(N\left(\frac{-y}{s}\right)+1)
\right\}\\
\nonumber
=
\frac{1}{2}\left(\frac{ms}{2\pi}\right)^2
\int_{1+L^{-1}/k}^\infty 
\frac{dz}{\sqrt{1-z^2}}
\int_{0}^{sk-sL^{-1}}
\frac{dy}{\sqrt{s^2k^2-y^2}}
\Bigr[\delta(\epsilon-y)
-
\delta(\epsilon+y)\Bigr]
\frac{1/2}{\mathrm{cosh}\left(\frac{x}{2T}\right)-\mathrm{cosh}\left(\frac{y}{2T}\right)}\cdot\frac{2e^{\frac{y}{2T}}}{e^{\frac{y}{T}}-1}.
\end{eqnarray}
This integral is over positive $y$, hence (as before) we use $\delta(\epsilon-y)-\delta(\epsilon+y)=\mathrm{sgn}(\epsilon)\delta(y-|\epsilon|)$. We find:
\begin{eqnarray}
F_2(\textbf{k},\epsilon)
=
-
\frac{\mathrm{sgn}(\epsilon)}{2}\left(\frac{ms}{2\pi}\right)^2
\frac{e^{\frac{|\epsilon|}{2T}}}{e^{\frac{|\epsilon|}{T}}-1}
\frac{\Theta[sk-|\epsilon|]}{\sqrt{s^2k^2-\epsilon^2}}
\int_{1+L^{-1}/k}^\infty 
\frac{dz}{\sqrt{z^2-1}}
\frac{1}{\mathrm{cosh}\left(\frac{|\epsilon|}{2T}\right)-\mathrm{cosh}\left(\frac{sk}{2T}z\right)}.
\end{eqnarray}

Let us now return to Eq.~\eqref{7}. We put $\varphi_\mathbf{p}=eE_xp_xB/m$ and then $\varphi_\mathbf{p}-\varphi_{\mathbf{p}+\mathbf{k}}=-eE_xk_xB/m$, where $k_x=k\cos(\beta+\phi)$. We find:
\begin{eqnarray}
\frac{eE_xp_0\cos(\beta)}{m}\frac{df_p}{d\epsilon_p}=-\sum_{\mathbf{k}}\int d\epsilon g_k^2\left(\frac{-eE_xk\cos(\beta+\phi)B}{Tm}\right)\left(-\epsilon\frac{df_p}{d\epsilon_p}\right)\delta(\epsilon_{\mathbf{p}+\mathbf{k}}-\epsilon_{\mathbf{p}}-\epsilon)F(\mathbf{k},\epsilon),
\end{eqnarray}
or cancelling out the matching terms,
\begin{eqnarray}
\label{Eq40}
p_0\cos(\beta)
=
-\frac{B}{T}\sum_{\mathbf{k}}g_k^2\int \epsilon d\epsilon \cos(\beta+\phi)
\delta(\epsilon_{\mathbf{p}+\mathbf{k}}-\epsilon_{\mathbf{p}}-\epsilon)F(\mathbf{k},\epsilon).
\end{eqnarray}
Since the function $F(\mathbf{k},\epsilon)$ depends on the absolute value $|\mathbf{k}|$, using
\begin{eqnarray}
\sum_{\mathbf{k}}=\int_0^\infty \frac{kdk}{2\pi}\int_0^{2\pi}\frac{d\phi}{2\pi},
\end{eqnarray}
we come to (denoting $v_0=p_0/m$)
\begin{eqnarray}
\int_0^{2\pi}\frac{d\phi}{2\pi}\cos(\beta+\phi)
\delta\left(\frac{p_0k}{m}\cos(\phi)+\frac{k^2}{2m}-\epsilon\right)
=
\cos(\beta)\int_0^{2\pi}\frac{d\phi}{2\pi}\cos(\phi)
\delta\left(v_0k\cos(\phi)+\frac{k^2}{2m}-\epsilon\right)
\\
\nonumber
=
\cos(\beta)
\left(\frac{\epsilon-k^2/(2m)}{v_0k}\right)
\frac{1}{\pi}
\frac{\Theta[v_0^2k^2-(\epsilon-k^2/(2m))^2]}{\sqrt{v_0^2k^2-(\epsilon-k^2/(2m))^2}}
\approx
\cos(\beta)
\left(\frac{\epsilon-k^2/(2m)}{v_0k}\right)
\frac{1}{\pi}
\frac{\Theta[v_0^2k^2-\epsilon^2]}{\sqrt{v_0^2k^2-\epsilon^2}}.
\end{eqnarray}
Substituting this result in Eq.~\eqref{Eq40} gives
\begin{eqnarray}
p_0\cos(\beta)
=
-\frac{B}{T}
\frac{\cos(\beta)}{2\pi^2}
\int_0^\infty \frac{k^2dkg_k^2}{v_0k}\int_{-\infty}^\infty \epsilon d\epsilon \left(\epsilon-k^2/(2m)\right)
\frac{\Theta[v_0^2k^2-\epsilon^2]}{\sqrt{v_0^2k^2-\epsilon^2}}
F(k,\epsilon).
\end{eqnarray}
Since $F(k,\epsilon)$ is an odd function due to the term $\mathrm{sgn}(\epsilon)$, we find:
\begin{eqnarray}
p_0
=
-\frac{B}{\pi^2T2mv_0}
\int_0^\infty k^3dkg_k\int_0^{v_0k} d\epsilon \frac{\epsilon F(k,\epsilon)}{\sqrt{v_0^2k^2-\epsilon^2}}.
\end{eqnarray}
It is convenient to introduce a new variable t: $\epsilon\rightarrow skt$, which yields:
\begin{eqnarray}
2\pi^2p_0^2T
&=&
s^2B
\int_0^\infty k^4dkg_k
\int_0^{v_0/s} 
\frac{tdt}{\sqrt{v_0^2-s^2t^2}}
F(k,skt)\\
\nonumber
&=&
s^2B
\int_0^\infty k^4dkg_k
\int_0^{v_0/s} 
\frac{tdt}{\sqrt{v_0^2-s^2t^2}}
\left(\frac{ms}{4\pi}\right)^2
\frac{1}{\sinh(\frac{sk}{2T}t)}
\left\{
\frac{\Theta(t-1)}{sk\sqrt{t^2-1}}
\int_0^1\frac{dz}{\sqrt{1-z^2}}
\frac{1}{\cosh(\frac{sk}{2T}t)-\cosh(\frac{sk}{2T}z)}\right.
\\
\nonumber
&&~~~~~~~~~~~~~~~~~~~~~~~~~~~~~~~~~~
\left.-
\frac{\Theta(1-t)}{sk\sqrt{1-t^2}}
\int_1^\infty \frac{dz}{\sqrt{z^2-1}}   \frac{1}{\cosh(\frac{sk}{2T}t)-\cosh(\frac{sk}{2T}z)}
\right\}.
\end{eqnarray}
Cancelling out $sk$, we get:
\begin{eqnarray}
2\pi^2p_0^2T
&=&
sB
\left(\frac{ms}{4\pi}\right)^2
\int_0^\infty k^3dkg^2_k
\int_0^{v_0/s} 
\frac{tdt}{\sqrt{v_0^2-s^2t^2}}
\frac{1}{\sinh(\frac{sk}{2T}t)}
\left\{
\frac{\Theta(t-1)}{\sqrt{t^2-1}}
\int_0^1\frac{dz}{\sqrt{1-z^2}}
\frac{1}{\cosh(\frac{sk}{2T}t)-\cosh(\frac{sk}{2T}z)}\right.
\\
\nonumber
&&~~~~~~~~~~~~~~~~~~~~~~~~~~~~~~~~~~
\left.-
\frac{\Theta(1-t)}{\sqrt{1-t^2}}
\int_1^\infty \frac{dz}{\sqrt{z^2-1}}   \frac{1}{\cosh(\frac{sk}{2T}t)-\cosh(\frac{sk}{2T}z)}
\right\}.
\end{eqnarray}

Now let us consider the integral
\begin{gather}\label{int.1}
J=\int_0^\infty k^3dkg^2_k
\int_0^{v_0/s}
\frac{tdt}{\sqrt{v_0^2-s^2t^2}}
\frac{1}{\sinh(\frac{sk}{2T}t)}\times\\\nonumber\times
\left\{
\frac{\Theta(t-1)}{\sqrt{t^2-1}}
\int_0^1\frac{dz}{\sqrt{1-z^2}}
\frac{1}{\cosh(\frac{sk}{2T}t)-\cosh(\frac{sk}{2T}z)}\right.
\left.-
\frac{\Theta(1-t)}{\sqrt{1-t^2}}
\int_1^\infty \frac{dz}{\sqrt{z^2-1}}   \frac{1}{\cosh(\frac{sk}{2T}t)-\cosh(\frac{sk}{2T}z)}
\right\}.
\end{gather}
We assume that $v_0>s$ (that is typical for real structures). Thus, we have to deal with the expression
\begin{eqnarray}
\label{int.2}
J&=&\int_0^\infty k^3dkg^2_k
\times\\
\nonumber
&&\times
\Biggl[
\int\limits_{1+L^{-1}/k}^{v_0/s}
\frac{tdt}{\sqrt{v_0^2-s^2t^2}}
\frac{1}{\sinh(\frac{sk}{2T}t)}\frac{1}{\sqrt{t^2-1}}
\int_0^{1-L^{-1}/k}\frac{dz}{\sqrt{1-z^2}}
\frac{1}{\cosh(\frac{sk}{2T}t)-\cosh(\frac{sk}{2T}z)}-\\
\nonumber
&&-
\int\limits_0^{1-L^{-1}/k}
\frac{tdt}{\sqrt{v_0^2-s^2t^2}}
\frac{1}{\sinh(\frac{sk}{2T}t)}\frac{1}{\sqrt{1-t^2}}
\int_{1+L^{-1}/k}^\infty \frac{dz}{\sqrt{z^2-1}}   \frac{1}{\cosh(\frac{sk}{2T}t)-\cosh(\frac{sk}{2T}z)}
\Biggr].
\end{eqnarray}
The two terms in the second and third lines diverge at $t\sim z\sim 1$. Thus we introduce new variables $t-1=u$ and $1-z=v$ in the first term and $1-t=u;\,z-1=v$ in the second one. The first term reads

\begin{eqnarray}\label{int.3}
\int\limits_{L^{-1}/k}^{v_0/s-1}
\frac{(1+u)du}{\sqrt{v_0^2-s^2(1+u)^2}}
\frac{1}{\sinh\left[\frac{sk}{2T}(1+u)\right]}\frac{1}{\sqrt{u(2+u)}}
\int_{L^{-1}/k}^1\frac{dv}{\sqrt{v(2-v)}}
\frac{1}{\cosh\left[\frac{sk}{2T}(1+u)\right]-\cosh\left[\frac{sk}{2T}(1-v)\right]},
\end{eqnarray}
and the second is
\begin{gather}\label{int.4}
\int\limits_{L^{-1}/k}^{1}
\frac{(1-u)du}{\sqrt{v_0^2-s^2(1-u)^2}}
\frac{1}{\sinh\left[\frac{sk}{2T}(1-u)\right]}\frac{1}{\sqrt{u(2-u)}}
\int\limits_{L^{-1}/k}^\infty\frac{dv}{\sqrt{v(2+v)}}
\frac{1}{\cosh\left[\frac{sk}{2T}(1-u)\right]-\cosh\left[\frac{sk}{2T}(1+v)\right]}.
\end{gather}
Now expanding these expressions for small $u$ and $v$, we find for the first term
\begin{gather}\label{int.5}
\frac{2T}{2sk\sqrt{v_0^2-s^2}}\frac{1}{\sinh^2\left[\frac{sk}{2T}\right]}\int\limits_{L^{-1}/k}^{v_0/s-1}
\frac{du}{\sqrt{u}}
\int_{L^{-1}/k}^1\frac{dv}{\sqrt{v}}
\frac{1}{u+v},
\end{gather}
and for the second term
\begin{gather}\label{int.6}
-\frac{2T}{2sk\sqrt{v_0^2-s^2}}\frac{1}{\sinh^2\left[\frac{sk}{2T}\right]}\int\limits_{L^{-1}/k}^{1}
\frac{du}{\sqrt{u}}
\int\limits_{L^{-1}/k}^\infty\frac{dv}{\sqrt{v}}
\frac{1}{u+v}.
\end{gather}
If $v_0\gg s$, we finally find
\begin{eqnarray}\label{int.7}
J&=&\frac{2T}{sv_0}\int\limits_{L^{-1}}^\infty\frac{k^2g_k^2dk}{\sinh^2\left[\frac{sk}{2T}\right]}\int\limits_{L^{-1}/k}^{1}
\frac{du}{\sqrt{u}}
\int\limits_{L^{-1}/k}^\infty\frac{dv}{\sqrt{v}}
\frac{1}{u+v}\\
&=&\frac{2\pi T}{sv_0}\int\limits_{L^{-1}}^\infty\frac{k^2g_k^2dk}{\sinh^2\left[\frac{sk}{2T}\right]}\ln(kL).
\end{eqnarray}
Here it becomes clear, why we had to introduce the cut-offs $sL^{-1}$. Otherwise the integrals over $u$ and $v$ in~\eqref{int.7} would be diverging like $\ln(1/0)$.

The resistivity, after restoring the constants, becomes
\begin{eqnarray}
\rho=\frac{ms^2}{32\pi^2e_0^2m\varepsilon_F^2}\int\limits_{L^{-1}}^\infty
\frac{k^2g_k^2dk}{\sinh^2\left[\frac{\hbar sk}{2k_BT}\right]}\ln(kL).
\end{eqnarray}

We can evaluate the closed form of the integral for the two limiting cases of low and high temperatures. First, we change the integration variable $x=2kl$ and write
\begin{eqnarray}
I\equiv\int_0^\infty\frac{k^2e^{-2kl}dk}{\sinh^2\left[\frac{\hbar sk}{2k_BT}\right]}\ln(kL)=\int_0^\infty\left(\frac{x}{2l}\right)^2\frac{e^{-x}\ln(\frac{Lx}{2l})}{\sinh^2\left(\frac{T_\textrm{BG}}{T}x\right)}dx.
\end{eqnarray}
For high temperatures $T\gg T_\textrm{BG}$,
\begin{eqnarray}
\sinh^2\left(\frac{T_\textrm{BG}}{T}x\right)\approx\left(\frac{T_\textrm{BG}}{T}x\right)^2,
\end{eqnarray}
and the fact that the main contribution of the integral comes from $0\leq x\lesssim 1$ gives
\begin{eqnarray}
I\approx\left(\frac{T}{T_{BG}}\right)^2\frac{1}{(2l)^3}\left[\ln\left(\frac{L}{2l}\right)-\gamma_C\right],
\end{eqnarray}
where $\gamma_C$ is the Euler gamma function.

For low temperatures $T\ll T_\textrm{BG}$, we obtain
\begin{eqnarray}
I\approx\left(\frac{T}{T_\textrm{BG}}\right)^3\frac{1}{(2l)^3}\ln\left(\frac{L}{2l}\right)\frac{\pi^2}{6}.
\end{eqnarray}


\section{Appendix C: Screening}
In this section, we calculate the screening factor $\epsilon_k$. In the presence of the condensate, it takes a usual form~\cite{Fetter}
\begin{eqnarray}
\epsilon_k=(1-v_k\Pi_k)(1-gP_k)-V_k^2\Pi_kP_k,
\end{eqnarray}
where $\Pi_k=-m/\pi$ and $P_k=-4Mn_c/k^2$ are the polarization operators for the electrons and exciton condensate, respectively, $v_k=2\pi e^2/k$ is the Coulomb interaction between electrons, $g=4\pi e^2d/\epsilon_0$, and $V_k=ge^{-kl}/2$ is the electron-exciton interaction.
After some algebra, we obtain
\begin{eqnarray}
\epsilon_k=1+\frac{2}{a_Bk}+\frac{1}{k^2\xi^2}+\frac{2}{a_Bk}\frac{1}{k^2\xi^2}\left(1-\frac{kd}{2}e^{-2kl}\right),
\end{eqnarray}
where $a_B$ is the Bohr radius.
For $l/d>1$ ,
\begin{eqnarray}
1-\frac{kd}{2}e^{-2kl}\approx 1,
\end{eqnarray}
hence we have
\begin{eqnarray}
\epsilon_k=\left(1+\frac{2}{a_Bk}\right)\left(1+\frac{1}{k^2\xi^2}\right).
\end{eqnarray}

To account for the screening in our calculation of resistivity in Appendices A and B, we should simply replace 
\begin{eqnarray}
|g_k|^2\rightarrow\left|\frac{g_k}{\epsilon_k} \right|^2.
\end{eqnarray}
%


\section{Appendix D: Phonon-assisted electron resistivity}
To calculate the phonon-limited electron resistivity, we use the electron-phonon interaction Hamiltonian
\begin{eqnarray}
H_{e-ph}=\frac{D}{L}\sum_{\mathbf{p}\mathbf{q}}\left(\frac{\hbar}{2\rho_d}\right)^{1/2}\frac{q}{\sqrt{\omega_q}}c^\dagger_{\mathbf{p}+\mathbf{q}}c_\mathbf{p}(a_\mathbf{q}+a^\dagger_{-\mathbf{q}}),
\end{eqnarray}
where $D$ is the deformation potential and $\rho_d$ is the ion density. The screening was calculated in~\cite{Kaasbjerg2013}. Here, we only present the results:
\begin{eqnarray}
\label{phscreen}
\epsilon(q,T,\mu)=1-\frac{e_0^2}{2\epsilon_0 q}\chi^0(q,T,\mu),
\end{eqnarray}
where
\begin{eqnarray}
\chi^0(q,T,\mu)=\int_0^\infty d\mu'\frac{\chi(q,0,\mu')}{4k_BT\cosh^2\frac{\mu-\mu'}{2k_BT}}
\end{eqnarray}
and $\chi(q,0,\mu')$ is the zero-temperature RPA polarizability. For $q\lessapprox 2k_F$, $\chi(q,0,\mu')=-\rho_\textrm{DOS}$, where $\rho_\textrm{DOS}$ is the density of states of the 2DEG. 

The calculation presented in Appendix A can be carried over, except that now we replace the matrix element by
\begin{eqnarray}
|M_q|^2=\frac{\hbar D^2q^2}{2\rho\omega_q},
\end{eqnarray}
and the bogolon sound velocity by the corresponding acoustic phonon sound velocity to find
\begin{eqnarray}
\rho=\frac{\pi\hbar^2}{e_0^2E_F}\frac{mD^2}{k_F^2\rho_d}\frac{(k_BT)^4}{(\hbar s)^5}\int_0^\infty\frac{du}{(2\pi)^2}\frac{u^4e^u}{(e^u-1)^2}(\Gamma_--\Gamma_+)|_{k_F}\frac{1}{\epsilon(u)}.
\end{eqnarray}
For the parameters that we are interested in, 
$\epsilon(u)$, given in Eq.~\eqref{phscreen}, is approximately equal to unity. That is, the screening is negligible.

\end{widetext}
%


\end{document}